\documentclass[aps,prd,nofootinbib]{revtex4}
\usepackage[utf8]{inputenc}
\usepackage[english]{babel}
\usepackage[T1]{fontenc}

\usepackage{calligra}
\usepackage{dutchcal} 

\usepackage{amsmath}
\usepackage{amsfonts}
\usepackage{amssymb}
\usepackage{graphicx}

\usepackage{lmodern}

\usepackage{hyperref}
\hypersetup{colorlinks, 
citecolor=blue,
linkcolor=red,
urlcolor=black}

\usepackage[normalem]{ulem}

\usepackage{booktabs}
\usepackage{ragged2e}

\begin{document}

\title{Quantum Brownian motion induced by a scalar field in Einstein's universe under Dirichlet and Neumann boundary conditions}
\author{E. J. B. Ferreira}
\email{ejbf@academico.ufpb.br}
\author{H. F. Santana Mota}
\email{hmota@fisica.ufpb.br}
\affiliation{Departamento de Física, Universidade Federal da Paraíba, Caixa Postal 5008, João Pessoa, Paraíba, Brazil}


\begin{abstract}
In this paper, the Quantum Brownian motion of a point particle induced by the quantum va\-cuum fluctuations of a real massless scalar field in Einstein's universe under Dirichlet and Neumann boundary conditions is studied. Using the Wightman functions, general expressions for the renormalized dispersion of the physical momentum are derived. Distinct expressions are found for the dispersion associated with each component of the particle's physical momentum, indicating that the global properties of homogeneity and isotropy of space are lost, as a consequence of the introduced boundary conditions. Divergences also arise and are related to the compact nature of Einstein's universe and the introduced boundary conditions.
\end{abstract}


\maketitle

\section{Introduction}

Induced quantum Brownian motion (IQBM) is one of the phenomena originated by vacuum fluctuations of quantum fields, similar to the very known Casimir effect \cite{casimir1948attraction}. The conceptual image of the physical phenomenon related to the IQBM corresponds to a point particle (also called test particle) interacting, for instance, with an electromagnetic or scalar quantum field. Hence, the quantum fluctuations of the vacuum state of the field make the test particle to follow a random trajectory.

This elementary system has been constantly investigated in recent years, considering distinct scenarios, both in electromagnetic \cite{yu2004vacuum,yu2004brownian,yu2006brownian,seriu2008switching,seriu2009smearing,de2016probing,
de2019remarks} and scalar field \cite{gour1999will,de2014quantum,camargo2018vacuum,camargo2019vacuum,Camargo:2020fxp,Ferreira:2023uxs,Ribeiro:2023yrt} models in flat spacetime. The consequences, for instance, of distinct topologies in the IQBM of particles in Minkowski spacetime have been studied by considering the vacuum fluctuations of the electromagnetic field \cite{Bessa:2019aar}. Furthermore,  in Refs. \cite{bessa2009brownian,bessa2017quantum,mota2020induced,ferreira2022quantum} the IQBM of particles in a flat Friedmann-Lemaitre-Robertson-Walker (FLRW) spacetime models has been also studied. 

In a recent paper \cite{Ferreira:2023gzv} we have investigated the IQBM in a curved spacetime with positive constant curvature, which defines Einstein's universe (a closed or compact spacetime model). Among other things, there we calculated the $\langle(\Delta\hat{\mathcal{p}}^{i})^{2}\rangle$ dispersions for the $\hat{\mathcal{p}}^{i}$ components of the physical momentum associated with the point particle and observe a homogeneous and isotropic result for the dispersions, which are typical properties of FLRW spacetime models. Periodic divergences were also found and explained as a consequence of the compact nature of Einstein's universe, as well as a possible consequence of the classical nature of the geometry adopted. Now, in the present paper, we wish to investigate the IQBM of a point particle in Einstein's universe under  Dirichlet and Neumann boundary conditions.

From a theoretical viewpoint, Einstein's universe is very interesting because it eliminates some intrinsic technical problems of quantum field theory in curved spacetimes. In fact, in curved backgrounds it is well known that the definition for the quantum vacuum state can be ambiguous and related to this ambiguity arise the phenomenon of particle production \cite{birrell1984quantum,Ford:2021syk}. For a recent review about particle production see Ref. \cite{Ford:2021syk}. However, Einstein's universe is static, so vacuum state definition is not ambiguous \cite{birrell1984quantum,Altaie:2001vv,Altaie:2002tv}. In addition, as argued in Ref. \cite{Altaie:2002tv}, it is a very important theoretical model that allows us to explore and study how the spacetime curvature can influence the phenomena arising in the context of quantum field theory.

Regarding the structure of this paper, in Section \ref{sec2}, first, we briefly define the spacetime geometry in which we will carry out the study. Next, we obtain the normalized solutions of the Klein-Gordon equation and the corresponding Wightman functions. In Section \ref{sec3}, we review the point particle dynamics and derive the equations of motion that will be used to study the momentum dispersion, which corresponds to the vacuum expectation value (VEV) of the particle's squared momentum. Finally, in Section \ref{sec4} we summarize the main results obtained throughout the paper.

\section{Normalizated solutions and Wightman functions}\label{sec2}

In order to maintain the text continuous and clear, in this section, first we review the obtaining of the normalized solutions of the Klein-Gordon equation in Einstein's universe and the corresponding Wightman function. Then, based on these discussion, we obtain the Wightman function for Einstein's universe under Dirichlet and Neumann boundary conditions. 
%
%

\subsection{Einstein's universe}\label{sec2A}
%

Now, we must briefly present the background geometry of the curved spacetime on which our investigation will be carry out, that is, the spacetime in which both the point particle and scalar field exist. One of the main elements of the modern Cosmology consist of the Friedmann-Lemaitre-Robertson-Walker (FLRW) metric tensor, which is defined by the line element \cite{adler2021general,d1992introducing,schutz2009first}
\begin{equation}\label{metricFLRW1}
ds^{2} = c^{2}dt^{2} - a^{2}(t)d\mathcal{r}^{2},
\end{equation}
with 
\begin{equation}\label{metricFLRW2}
d\mathcal{r}^{2} = \left\{\frac{dr^{2}}{1-kr^{2}}+r^{2}[d\theta^{2}+\sin^{2}(\theta)d\phi^{2}]\right\}.
\end{equation}
In Eq. \eqref{metricFLRW1} $a(t)$ is the scale factor and in Eq. \eqref{metricFLRW2} the constant parameter $k$, associated with the spacial geometry, can take on three specific values, namely, $k=(-1,0,+1)$. Also, as we know, these three values of the parameter $k$ describe, respectively, an open, flat and closed universe, but all equally homogeneous e isotropic \cite{adler2021general,schutz2009first}. From now on we will use units such that $c=\hbar=1$.

Einstein's universe corresponds to the particular case $k=+1$ of the FLRW line element with a constant scale factor, namely, \cite{schutz2009first,d1992introducing,islan2004}
\begin{eqnarray}\label{metricEinsteinUniverse}
ds_{\text{EU}}^{2} = dt^{2}-a^{2}_{0}\left\{d\chi^{2}+\sin^{2}(\chi)\left[d\theta^{2}+\sin^{2}(\theta)d\phi^{2}\right]\right\},
\end{eqnarray}
%
%
where in order to establish the above equation we have introduced the new coordinate $r=\sin(\chi)$ in Eq. \eqref{metricFLRW2}. The line element $ds_{\text{EU}}^{2}$ describes a static universe model, whose spatial section of spacetime is closed or compacted, and of finite volume. 

In Eq. \eqref{metricEinsteinUniverse}, $a_{0} = a(t=t_{0})$ and corresponds to a constant value for the scale factor at an arbitrary instant of time $t=t_{0}$, which defines a hypersurface $t=$ constant in this spacetime. In addition, by geometric reasons, $a_{0}$ is commonly known as the radius of Einstein's universe. Other geometric properties of this spacetime can be consulted, for instance, in Refs. \cite{d1992introducing} and \cite{islan2004}. The validity range of the coordinates in $ds_{\text{EU}}^{2}$ are such that $0\leq t\leq\infty$, $0\leq\chi\leq\pi$, $0\leq\theta\leq\pi$ and $0\leq\phi\leq 2\pi$. It is observed that the angular sector $(\theta,\phi)$ corresponds to the usual angles of the spherical coordinates. Thus, as we shall see, the part of the solution of the Klein-Gordon equation referring to these angles corresponds to the spherical harmonics.

\subsection{Solution of the Klein-Gordon equation}\label{sec2B}

%
According to the formalism of quantum field theory, the solution of the Klein-Gordon equation is an essential element  for the construction of the scalar field operator $\hat{\psi}(x)$. In general, the dynamics of a free scalar field $\psi(x)$, with mass $m_{\textrm{F}}$, propagating through an arbitrary spacetime, is given by the expression \cite{birrell1984quantum,fulling1989aspects}
\begin{equation}\label{GenExpEQKG1}
\left(\Box+m_{\textrm{F}}^{2}+\xi R\right)\psi(x)=0,
\end{equation}
where the second order covariant differential operator $\Box\psi = \nabla_{\mu}\nabla^{\mu}\psi$ is given by 
\begin{equation}\label{GenExpEQKG2}
\Box\psi(x) = \frac{1}{\sqrt{-g}}\partial_{\mu}[\sqrt{-g}g^{\mu\nu}\partial_{\nu}\psi(x)],
\end{equation}
with the determinant $g=\textrm{Det}(g_{\mu\nu})$. In the present case, remembering that $ds^{2}=g_{\mu\nu}dx^{\mu}dx^{\nu}$, the metric tensor $g_{\mu\nu}$ can be obtained directly from Eq. \eqref{metricEinsteinUniverse} so that $g=-a_{0}^{6}\sin^{4}(\chi)\sin^{2}(\theta)$.
%

%
In Eq. \eqref{GenExpEQKG1} the term $\xi R\psi$ represents a contribution that comes from the coupling between the scalar field $\psi$ and the effects of gravity, where the coupling strength is defined by the constant $\xi$. The Ricci scalar $R$, in which the effects of gravity are encoded, can be calculated by contracting the Ricci tensor $R_{\mu\nu}$ according to the relation $R = g^{\mu\nu}R_{\mu\nu}$, with \cite{adler2021general}
\begin{eqnarray}\label{RicciTensor}
R_{\mu\nu}(x) = \Gamma^{\beta}_{\beta\nu,\mu}-\Gamma^{\beta}_{\mu\nu,\beta}+\Gamma^{\beta}_{\alpha\mu}\Gamma^{\alpha}_{\beta\nu}-\Gamma^{\beta}_{\alpha\beta}\Gamma^{\alpha}_{\mu\nu},
\end{eqnarray}
where and Christoffel symbols are given by
\begin{eqnarray}\label{ChristoffelSymbols}
\Gamma^{\alpha}_{\mu\nu}(x) = \frac{1}{2}g^{\alpha\beta}\left( g_{\beta\mu,\nu}+g_{\beta\nu,\mu}-g_{\mu\nu,\beta}\right).
\end{eqnarray}
%
%
In addition, regarding the coupling constant in Eq. \eqref{GenExpEQKG1}, specifically, the cases $\xi=0$ and $\xi\neq 0$ represent the minimum and non-minimum coupling cases, respectively. If, on the other hand, one assumes the particular value $\xi(n) = \frac{(n-2)}{4(n-1)}$, we have the case known as conformally coupled, where $n$ corresponds to the number of spacetime dimensions \cite{birrell1984quantum}. As we are considering a four-dimensional spacetime $n=4$ and, consequently, $\xi=\frac{1}{6}$. It is important to note that even in the case of minimal coupling $(\xi=0)$, the field still sees a curved spacetime because of the geometrical informations contained in $\Box\psi(x)$ term, as we can see from Eq. \eqref{GenExpEQKG2}.

%
Assuming that the field $\psi(x)$ can be decomposed as the product of independent and separable solutions $\psi(t,\chi,\theta,\phi)=T(t)\mathcal{R}(\chi)\Theta(\theta)\Phi(\phi)$, we can show that a set of solutions of Eq. \eqref{GenExpEQKG1}, representing the positive frequency modes of the scalar field, is given by \cite{ford1975quantum,ford1976quantum,Ferreira:2023gzv}
\begin{eqnarray}\label{GEModes}
\psi_{\sigma}(t,\chi,\theta,\phi) = \textrm{N}\sin^{\ell}(\chi)C_{n-\ell}^{\ell +1}(\cos\chi)Y_{\ell}^{m}(\theta,\phi)e^{-i\omega_{n} t},
\end{eqnarray}
%
%
where $C^{\alpha}_{m}(z)$ are known as Gegenbauer polynomials or ultraspherical polynomials \cite{abramowitz1970handbook,gradshtein2007} and $Y_{\ell}^{m}(\theta,\phi)$ are the spherical harmonics \cite{arfken2005mathematical}, with $n=0,1,2,\ldots$, $\ell =0,1,2,\ldots, n$ e $-\ell\leq m\leq \ell$. In Eq. \eqref{GEModes} the eigenfrequencies $\omega_{n}$ are defined by the expression
\begin{eqnarray}\label{freqOfModes}
\omega_{n} = \left[ \dfrac{n(n+2)}{a_{0}^{2}} + M^{2} \right]^{1/2},
\end{eqnarray}
%
%
with $M^{2} = m_{F}^{2} + \xi R$ and the subscript $\sigma = (n,\ell,m)$ stands for the set of numbers responsible for specifying each of the field's modes.  In addition, as we shall see below, the constant N in Eq. \eqref{GEModes} can be obtained, so that we can construct a set of orthonormal solutions.

%
At this point it is important to note that, in principle, $R$ can depend on the spacetime coordinates, as we can see from Eqs. \eqref{RicciTensor} and \eqref{ChristoffelSymbols}. Thus, in this case, its structure can significantly modify the solutions of the Klein-Gordon equation. However, in Einstein's universe the Ricci scalar is constant, namely, $R=6a_{0}^{-2}$, so that we can identify the constant effective mass $M^2=m^2 + \frac{6}{a_0^2}$ in Eq. \eqref{freqOfModes}. Next we will establish the methodology necessary to obtain the Wightman function which is a fundamental quantity in our studies.

\subsection{Quantum scalar field and Wightman functions}\label{sec2C}

%
To study the IQBM of a point particle by a scalar field in Einstein's universe with Dirichlet and Neumann boundary conditions we need to find the corresponding Wightman functions. After obtaining the solutions $\psi_{\sigma}(x)$ of the Klein-Gordon equation \eqref{GenExpEQKG1}, which correspond to the modes of the scalar field \eqref{GEModes}, we can normalize these modes and use them to construct the field operator $\hat{\psi}(x)$ through the following expression \cite{birrell1984quantum}:
\begin{equation}\label{GEFieldOperator}
\hat{\psi}(x) = \sum_{\sigma}\left[\hat{a}_{\sigma}\psi_{\sigma}(x)+\hat{a}_{\sigma}^{\dagger}\psi_{\sigma}^{*}(x) \right],
\end{equation}
%
%
where the coefficients $\hat{a}_{\sigma}$ and $\hat{a}_{\sigma}^{\dagger}$ are the creation and annihilation operatos, respectively. These operators, which does not depend on the spacetime coordinates, create and annihilate the quanta associated with the  field $\hat{\psi}(x)$ and satisfy the fundamental property of commutation
\begin{equation}\label{Sec2RCommutation}
[\hat{a}_{\sigma},\hat{a}_{\sigma'}^{\dagger}]=\delta_{\sigma\sigma'}
\end{equation}
%
%
and the elementary algebra of the vacuum state
\begin{subequations}\label{Sec2VacuumAlgebra}
\begin{equation}\label{Sec2VacuumAlgebra1}
\hat{a}_{\sigma}|0\rangle = 0\ \ (\ \forall\ \sigma\ ),
\end{equation}
%
%
and
\begin{equation}\label{Sec2VacuumAlgebra3}
\langle 0|\hat{a}_{\sigma}\hat{a}_{\sigma'}^{\dagger}|0\rangle = \delta_{\sigma\sigma'}.
\end{equation}
\end{subequations} 
%
%
As we have mentioned before, in Eqs. \eqref{GEFieldOperator}, \eqref{Sec2RCommutation} and \eqref{Sec2VacuumAlgebra}, the subindex $\sigma$ represents in a compacted form the set of values for quantum numbers related to the modes of the scalar field, which can be either discrete or continuous. However, in the present case, we work with the set of discrete indices $\sigma=(n,\ell,m)$, as already pointed out. Then, the sum symbol in Eq. \eqref{GEFieldOperator} stands for three discrete sums in the respective range of the quantum numbers.

In curved spacetime, the normalization condition that we must use for normalize the modes $\psi_{\sigma}(x)$ in Eq. \eqref{GEFieldOperator} is defined by the relation
\begin{eqnarray}\label{GEnormalization}
-i\int dx^{3}\sqrt{-g}[\psi_{\sigma}(\partial_{t}\psi^{*}_{\sigma'})-(\partial_{t}\psi_{\sigma})\psi^{*}_{\sigma'}]=\delta_{\sigma\sigma'},
\end{eqnarray}
%
%
which by definition corresponds to the scalar product between the modes $\psi_{\sigma}(x)$ e $\psi_{\sigma'}(x)$, symbolically represented by the notation $(\psi_{\sigma},\psi_{\sigma'})$. This relation allow us to construct a set of orthonormal solutions, i.e., which obey both orthogonality and normality conditions, mathematically expressed by properties $(\psi_{\sigma},\psi_{\sigma'})= (\psi_{\sigma},\psi_{\sigma'}^{*})=0$ and $(\psi_{\sigma},\psi_{\sigma'})=-(\psi_{\sigma}^{*},\psi_{\sigma'}^{*})=\delta_{\sigma\sigma'}$ \cite{birrell1984quantum}.

%
Considering the previous discussions, from Eqs. \eqref{GEFieldOperator} and \eqref{GEModes}, now we can obtain the positive frequency Wightman function (PFWF) for the scalar field (duly normalized) in Einstein's universe using the general expresion
\begin{eqnarray}\label{GEwightmanfunction}
\textrm{W}(x,x')&=&\langle 0|\hat{\psi}(x)\hat{\psi}(x') |0\rangle \nonumber \\
&=& \sum_{\sigma}\psi_{\sigma}(x)\psi_{\sigma}^{*}(x').
\end{eqnarray}
%
%
To establish the second line of the above equation we have used the algebras of the creation and annihilation operators in the vacuum state shown in Eqs. \eqref{Sec2VacuumAlgebra}.
%

Next, using the methodology and expressions \sout{in} {\color{blue}of} this section, we present a derivation for the Wightman functions necessary to study the IQBM in Einstein's universe under Dirichlet and Neumann boundary conditions. In order to clarify the discussion, we would like to present the outline of the algorithm that we will be following in subsequent subsections. In summary, the methodology consists of first obtaining the modes, solving Eq. \eqref{GenExpEQKG1}, and normalizing them, using Eq. \eqref{GEnormalization}. Then, we use Eq. \eqref{GEFieldOperator} to construct the field operator and, finally, Eq. \eqref{GEwightmanfunction} to find the Wightman functions.

\subsubsection{PFWF for the Einstein's universe}\label{sec2C1}

%
For the case of Einstein's universe (E) without boundary conditions, from Eqs. \eqref{GEModes} and \eqref{GEnormalization}, after some computations and simplifications, we find that
\begin{eqnarray}\label{Sec2eq00a}
\psi_{\sigma}^{\textrm{(E)}}(t,\chi,\theta,\phi) = \textrm{N}_{n\ell}^{\textrm{(E)}}\sin^{\ell}(\chi)C_{n-\ell}^{\ell +1}(\cos(\chi))Y_{\ell}^{m}(\theta,\phi)e^{-i\omega_{n} t}
\end{eqnarray}
are the normalized modes, with
\begin{eqnarray}\label{Sec2eq00b}
\textrm{N}_{n\ell}^{\textrm{(E)}}=\left\{ \dfrac{2^{2\ell}(n+1)(n-\ell)![\Gamma(\ell +1)]^{2}}{\pi a_{0}^{3}\omega_{n}\Gamma(\ell +n +2)}\right\}^{1/2}.
\end{eqnarray}
%
%
It is worth pointing out that, in the above equations, superscript E indicates that the respective quantities refer to Einstein's universe without boundary conditions. In addition, to establish the normalization constant $\textrm{N}_{n\ell}^{\textrm{(E)}}$ we use the orthogonality relations of spherical harmonics \cite{arfken2005mathematical}, 
\begin{eqnarray}\label{prop_sphe_har_ort}
\int_{0}^{2\pi}d\phi\int_{0}^{\pi}d\theta\sin(\theta)[Y_{\ell}^{m}(\theta,\phi)]^{*}Y_{\ell'}^{m'}(\theta,\phi) = \delta_{\ell\ell'}\delta_{mm'},
\end{eqnarray}
and Gegenbauer polynomials \cite{gradshtein2007},
\begin{eqnarray}\label{prop_gegenb_ort}
\int_{-1}^{1}dz(1-z^2)^{\lambda -1/2}C^{\lambda}_{j}(z)C^{\lambda}_{k}(z) = \left\{
	\begin{array}{ll}
		&0\  \textrm{if} \ j\neq k, \\
		& \dfrac{\pi 2^{1-2\lambda}\Gamma(j+2\lambda)}{j!(\lambda + j)[\Gamma(\lambda)]^{2}}, \textrm{if} \ j=k \ (\lambda\neq 0).
	\end{array}\right.
\end{eqnarray}

%
According to the previous discussions, to obtain the Wightman function, we must now replace Eq. \eqref{Sec2eq00a} in Eq. \eqref{GEwightmanfunction}, with the sum symbol
\begin{eqnarray}\label{Sec2eq01}
\sum_{\sigma}\equiv \sum_{n=0}^{\infty}\sum_{\ell=0}^{n}\sum_{m=-\ell}^{\ell},
\end{eqnarray}
so that we find
\begin{eqnarray}\label{Sec2eq02}
\textrm{W}^{\textrm{(E)}}=\dfrac{1}{4\pi} \sum_{n=0}^{\infty}\sum_{\ell=0}^{n}\left|\textrm{N}_{n\ell}^{\textrm{(E)}}\right|^{2}(2\ell+1)e^{-i\omega_{n}\Delta t}\sin^{\ell}(\chi)\sin^{\ell}(\chi')C_{n-\ell}^{\ell +1}(\cos\chi)C_{n-\ell}^{\ell +1}(\cos\chi')P_{\ell}(\cos\gamma),
\end{eqnarray}
where we define $\Delta t = (t-t')$. To perform the sum in the discrete index $m$ we use the addition theorem of the spherical harmonics, which is given by relation \cite{arfken2005mathematical}
\begin{eqnarray}\label{prop_add_theo_sphe_harm1}
P_{\ell}(\cos(\gamma)) = \dfrac{4\pi}{2\ell +1}\sum_{m=-\ell}^{\ell}Y_{\ell}^{m}(\theta,\phi)[Y_{\ell}^{m}(\theta',\phi')]^{*},
\end{eqnarray}
with
\begin{eqnarray}\label{prop_add_theo_sphe_harm2}
\cos(\gamma) = \cos(\theta)\cos(\theta') + \sin(\theta)\sin(\theta')\cos(\phi - \phi').
\end{eqnarray}

%
In a superficial way, two crucial identities for the development of Eq. \eqref{Sec2eq02} up to its final form are the Gegenbauer polynomial addition theorem \cite{gradshtein2007} and the Abel-Plana formula \cite{Saharian:2007ph}. Given that only the structure of Eq. \eqref{Sec2eq02} and its final form (which we show below) are useful for the subsequent discussions, here, the formal details of this development are omitted. However, we recommend that the reader consult Ref. \cite{Ferreira:2023gzv} and the sources cited therein for the necessary details. Then, a possible representation of Eq. \eqref{Sec2eq02} is such that
\begin{eqnarray}\label{eqWF_EinsteinUniverseA}
\textrm{W}^{\textrm{(E)}}(x,x') = \dfrac{im_{F}}{8\pi a_{0}\sin\left(\frac{\Delta s}{a_{0}}\right)}\sum_{n=-\infty}^{\infty}\dfrac{(\Delta s +2\pi a_{0} n)}{\sigma_{n}}H_{1}^{(2)}(m_{F}\sigma_{n})
\end{eqnarray}
%
%
for the massive scalar field and \cite{dowker1976covariant}
\begin{eqnarray}\label{eqWF_EinsteinUniverseB}
\textrm{W}^{\textrm{(E)}}(x,x') = - \dfrac{1}{4a_{0}\pi^{2}}\sum_{n=-\infty}^{\infty}\dfrac{(\Delta s +2\pi a_{0}n)}{\sin\left(\frac{\Delta s}{a_{0}}\right)\sigma_{n}^{2}},
\end{eqnarray}
%
%
for the massless scalar field, with
\begin{eqnarray}\label{eqWF_EinsteinUniverseC}
\sigma_{n}^{2} = \Delta t^{2} - (\Delta s+2\pi a_{0} n)^{2}.
\end{eqnarray}
%
%
In Eq. \eqref{eqWF_EinsteinUniverseA} $H_{1}^{(2)}(z)$ is the Hankel function or Bessel function of the third kind \cite{gradshtein2007}. As we know, in the above equations $a_{0}$ is the radius of Einstein's universe, $\sigma_{n}$ represents the modulus of the spacetime separation vector and $\Delta s$ corresponds to the spatial separation between two vectors defined by the set of coordinates $(\chi,\theta,\phi)$ and $(\chi',\theta',\phi')$, respectively, whose internal angle $\alpha$ of the separtion is such that  \cite{Ozcan:2006jn,Ozcan:2001cr}
\begin{eqnarray}\label{Sec2eq03}
\cos(\alpha) = \cos(\chi)\cos(\chi')+\sin(\chi)\sin(\chi')\cos(\gamma),
\end{eqnarray}
%
%
where $\cos(\gamma)$ is given by Eq. \eqref{prop_add_theo_sphe_harm2}. In addition, in this spacetime the spatial distance $\Delta s$ and angle $\alpha$ are related by the expression $\Delta s=a_{0}\alpha$. Finally, we emphasize that Eqs. \eqref{eqWF_EinsteinUniverseA} and \eqref{eqWF_EinsteinUniverseB} correspond to the PFWF expressions for the massive and massless scalar fields in Einstein's universe.

\subsubsection{PFWF for Einstein's universe under Dirichlet and Neumann boundary conditions}\label{sec2C2}

In this part we will consider the Einstein's universe with the effect of Dirichlet and Neumann boundary conditions imposed on the scalar field, with the spacetime geometry determined by the line element \eqref{metricEinsteinUniverse} and the angular variable $\chi$ defined by the modified range $\chi=\left[0,\frac{\pi}{2}\right]$. It is noteworthy that, in Einstein's universe (without boundary conditions), previously discussed, the natural range of the $\chi$ coordinate is $\chi=[0,\pi]$. In the present case, we will use the Dirichlet and Neumann boundary conditions when the angular variable $\chi$ assumes the particular value $\chi=\frac{\pi}{2}$. For the Dirichlet (D) boundary condition, the modes must satisfy the condition
\begin{eqnarray}\label{dirichet_bc}
\left.\psi \right|_{\chi = \frac{\pi}{2}} = 0,
\end{eqnarray}
whereas for the Neumann (N) boundary condition
\begin{eqnarray}\label{neumann_bc}
\left.\left(\frac{\partial\psi}{\partial\chi}\right)\right|_{\chi=\frac{\pi}{2}} = 0.
\end{eqnarray}

From Eq. \eqref{GEModes}, we find that the normalized solutions that satisfy the relations \eqref{dirichet_bc} and \eqref{neumann_bc} are given by
\begin{eqnarray}\label{modesD&N}
\psi_{\sigma}^{\textrm{(i)}}(t,\chi,\theta,\phi) = \textrm{N}_{n\ell}^{\textrm{(i)}}\sin^{\ell}(\chi)C_{n-\ell}^{\ell +1}(\cos(\chi))Y_{\ell}^{m}(\theta,\phi)e^{-i\omega_{n} t},
\end{eqnarray}
where
\begin{eqnarray}\label{constnormD&N}
\textrm{N}_{n\ell}^{\textrm{(i)}}=\left\{ \dfrac{2^{2\ell+1}(n+1)(n-\ell)![\Gamma(\ell +1)]^{2}}{\pi a_{0}^{3}\omega_{n}\Gamma(\ell +n +2)}\right\}^{1/2},
\end{eqnarray}	
with i = (D, N). The normalization constant $\textrm{N}_{n\ell}^{\textrm{(i)}}$ was obtained through condition \eqref{GEnormalization} and using the orthogonality relations \eqref{prop_sphe_har_ort} and \eqref{prop_gegenb_ort}. In addition, the eigenfrequencies $\omega_{n}$ are defined in Eq. \eqref{freqOfModes}. The use of boundary conditions affects both the normalization constant and the values of the quantum numbers. In fact, comparing Eqs. \eqref{Sec2eq00b} and \eqref{constnormD&N} we note that $\textrm{N}_{n\ell}^{\textrm{(D,N)}} = \sqrt{2}\textrm{N}_{n\ell}^{\textrm{(E)}}$. Furthermore, observing the property of Gegenbauer polynomial $C_{n}^{\lambda}(0) = 0$, for $\lambda\neq 0$ and $n=2j+1$, with integer $j\geq 0$ \cite{abramowitz1970handbook}, in the Dirichlet case the range of quantum numbers is such that $n = \{1,2,3,\ldots\}$ and $\ell = \{0,1,2,\ldots,n\}$ with the constraint $(n-\ell)=\{1,3,5,\ldots\}$ (odd numbers). On the other hand, in the Neumann case the quantum numbers are such that $n = \{0,1,2,3,\ldots\}$ and $\ell = \{0,1,2,\ldots,n\}$ with the constraint $(n-\ell)=\{0,2,4,\ldots\}$ (even numbers). 

To obtain the corresponding Wightman functions for the Dirichlet and Neumann boundary conditions for the Einstein universe we use Eqs. \eqref{GEwightmanfunction} and \eqref{modesD&N}, so that we find
\begin{eqnarray}\label{wightman_fucntion_FDN}
\textrm{W}^{\textrm{(i)}} = {\sum_{n,\ell}}^{\textrm{(i)}}\textrm{W}_{n\ell}^{\textrm{(i)}},
\end{eqnarray}
where $i =$ (D, N) and we conveniently define
\begin{eqnarray}\label{coef_wfFDN}
\textrm{W}_{n\ell}^{\textrm{(i)}} = \frac{1}{4\pi}\left|\textrm{N}_{n\ell}^{\textrm{(i)}}\right|^{2}(2\ell+1)e^{-i\omega_{n}\Delta t} \sin^{\ell}(\chi)\sin^{\ell}(\chi')C_{n-\ell}^{\ell+1}(\cos(\chi))C_{n-\ell}^{\ell+1}(\cos(\chi'))P_{\ell}(\cos(\gamma)).
\end{eqnarray}
It is noteworthy that in the above relation $\cos(\gamma)$ is defined in Eq. \eqref{prop_add_theo_sphe_harm2}. The upper index in the sum symbol specifies the boundary condition and its respective ranges with the constraint of possibles values for the quantum numbers. Specifically, we have that for the Dirichlet case
\begin{eqnarray}\label{sum_symb_dirichlet}
{\sum_{n,\ell}}^{\textrm{(D)}}\textrm{W}_{n\ell}^{\textrm{(D)}}& \equiv & \sum_{n=1}^{\infty}\sum_{\ell=0}^{n}\textrm{W}_{n\ell}^{\textrm{(D)}}, \ \textrm{with}\ (n-\ell)=1,3,5,\ldots,
\end{eqnarray}
and for the Neumann case
\begin{eqnarray}\label{sum_symb_Neumann}
{\sum_{n,\ell}}^{\textrm{(N)}}\textrm{W}_{n\ell}^{\textrm{(N)}}& \equiv & \sum_{n=0}^{\infty}\sum_{\ell=0}^{n}\textrm{W}_{n\ell}^{\textrm{(N)}}, \ \textrm{with}\ (n-\ell)=0,2,4,\ldots.
\end{eqnarray}
From Eqs. \eqref{coef_wfFDN}, \eqref{Sec2eq02}, \eqref{Sec2eq00b} and \eqref{constnormD&N} we can notice that there is a relation between the coefficients of the Einstein's universe, $\textrm{W}_{n\ell}^{\textrm{(E)}}$, and the Einstein's universe under Dirichlet and Neumann boundary conditions, $\textrm{W}_{n\ell}^{\textrm{(D,N)}}$, namely, $\textrm{W}_{n\ell}^{\textrm{(D,N)}} = 2\textrm{W}_{n\ell}^{\textrm{(E)}}$. The $\textrm{W}_{n\ell}^{\textrm{(E)}}$ coefficients have the same mathematical structure as in Eq. \eqref{coef_wfFDN}, except for the constraints on the $n$ and $\ell$ indices.
This is a direct consequence of the normalization process. In fact, in Einstein's universe the normalization is performed with respect to the whole space. On the other hand, in Einstein's universe with the boundary conditions considered the normalization is performed considering a subspace such that $0\leq\chi\leq\frac{\pi}{2}$, which corresponds to half of the usual space, that is, without the boundary conditions. Then, this gives rise to the factor of 2 observed in the previous relations. For the derivation of the Wightman function presented below we based our analysis on Refs. \cite{Ozcan:2001cr} and \cite{Bayin:1997nj}.

Before proceeding, it is crucial for the following discussions to observe the behavior of the elements of the double series in Eq. \eqref{wightman_fucntion_FDN}. This is shown in Table \ref{Tab0}, in which, for understanding purposes, we have brought together the three subtables containing the W$^{\textrm{(E,D,N)}}_{n\ell}$ coefficients for each of the three cases discussed in this article. Table \ref{Tab0}-(a) shows the first terms for the case of Einstein's universe, Eq. \eqref{Sec2eq02}, in which no boundary conditions are imposed on the modes. Similarly, in Tables \ref{Tab0}-(b) and \ref{Tab0}-(c) we have the first possible terms of the series \eqref{wightman_fucntion_FDN} for the cases of Dirichlet and Neumann boundary conditions, so that $(n-\ell)$ must be odd and even, repectively.
\begin{table}[h!]
%
\caption{First values for the W$^{\textrm{(i)}}_{n\ell}$ coefficients of the Wightman functions referring to the cases of Einstein's universe without boundary conditions and of the Einstein's universe under Dirichlet and Neumann boundary conditions.}\label{Tab0}
\vspace{0.25cm}
\begin{minipage}[c]{0.45\linewidth}
\begin{center}
(a) Einstein's universe.
\end{center}
	\begin{tabular}{cccccccc}\toprule
	 	$n \diagdown \ell$ & 0 & 1 & 2 & 3 & 4 & 5 & $\cdots$ \\ \midrule\midrule
	  	0 & W$_{00}^{\textrm{(E)}}$ & & & & & & \\ \midrule
	  	1 & W$_{10}^{\textrm{(E)}}$ & W$_{11}^{\textrm{(E)}}$ & & & & & \\ \midrule
	  	2 & W$_{20}^{\textrm{(E)}}$ & W$_{21}^{\textrm{(E)}}$ & W$_{22}^{\textrm{(E)}}$ & & & & \\ \midrule
	  	3 & W$_{30}^{\textrm{(E)}}$ & W$_{31}^{\textrm{(E)}}$ & W$_{32}^{\textrm{(E)}}$ & W$_{33}^{\textrm{(E)}}$ & & & \\ \midrule
	  	4 & W$_{40}^{\textrm{(E)}}$ & W$_{41}^{\textrm{(E)}}$ & W$_{42}^{\textrm{(E)}}$ & W$_{43}^{\textrm{(E)}}$ & W$_{44}^{\textrm{(E)}}$ & & \\ \midrule
	  	5 & W$_{50}^{\textrm{(E)}}$ & W$_{51}^{\textrm{(E)}}$ & W$_{52}^{\textrm{(E)}}$ & W$_{53}^{\textrm{(E)}}$ & W$_{54}^{\textrm{(E)}}$ & W$_{55}^{\textrm{(E)}}$ & \\ \midrule
	  	$\vdots$ &   &   &   &   &   &   & $\ddots$ \\
	\end{tabular}
\end{minipage}
\vspace{0.5cm}

\begin{minipage}[c]{0.45\linewidth}
\begin{center}
(b) Einstein's universe: Dirichlet case.
\end{center}
	\begin{tabular}{cccccccc}\toprule
	 	$n \diagdown\ell$ & 0 & 1 & 2 & 3 & 4 & 5 & $\cdots$ \\ \midrule\midrule
	  	0 & (N) & & & & & & \\ \midrule
	  	1 & W$_{10}^{\textrm{(D)}}$ & (N) & & & & & \\ \midrule
	  	2 & (N) & W$_{21}^{\textrm{(D)}}$ & (N) & & & & \\ \midrule
	  	3 & W$_{30}^{\textrm{(D)}}$ & (N) & W$_{32}^{\textrm{(D)}}$ & (N) & & & \\ \midrule
	  	4 & (N) & W$_{41}^{\textrm{(D)}}$ & (N) & W$_{43}^{\textrm{(D)}}$ & (N) & & \\ \midrule
	  	5 & W$_{50}^{\textrm{(D)}}$ & (N) & W$_{52}^{\textrm{(D)}}$ & (N) & W$_{54}^{\textrm{(D)}}$ & (N) & \\ \midrule
	  	$\vdots$ &   &   &   &   &   &   & $\ddots$ \\ 
	\end{tabular}
\end{minipage}
\begin{minipage}[c]{0.45\linewidth}
\begin{center}
(c) Einstein's universe: Neumann case.
\end{center}
%
	\begin{tabular}{cccccccc}\toprule
	 	$n\diagdown \ell$ & 0 & 1 & 2 & 3 & 4 & 5 & $\cdots$ \\ \midrule\midrule
	  	0 & W$_{00}^{\textrm{(N)}}$ & & & & & & \\ \midrule
	  	1 & (D) & W$_{11}^{\textrm{(N)}}$ & & & & & \\ \midrule
	  	2 & W$_{20}^{\textrm{(N)}}$ & (D) & W$_{22}^{\textrm{(N)}}$ & & & & \\ \midrule
	  	3 & (D) & W$_{31}^{\textrm{(N)}}$ & (D) & W$_{33}^{\textrm{(N)}}$ & & & \\ \midrule
	  	4 & W$_{40}^{\textrm{(N)}}$ & (D) & W$_{42}^{\textrm{(N)}}$ & (D) & W$_{44}^{\textrm{(N)}}$ & & \\ \midrule
	  	5 & (D) & W$_{51}^{\textrm{(N)}}$ & (D) & W$_{53}^{\textrm{(N)}}$ & (D) & W$_{55}^{\textrm{(N)}}$ & \\ \midrule
	  	$\vdots$ &   &   &   &   &   &   & $\ddots$ \\ 
	\end{tabular}
\end{minipage}
\vspace{0.25cm}
\justifying 
%
%

\noindent Legend: In the tables above, the quantum numbers $n$ and $\ell\leq n$ identify the rows and columns, respectively. In table (a) there are no constrains for the values of $n$ and $\ell$, so they can assume any values in their respective ranges. In the cases of tables (b) and (c), the values of $n$ and $\ell$ are subject to the constraint $(n-\ell)=\{1,3,5,\ldots\}$, for the Dirichlet condition, and $(n-\ell)=\{0,2,4,\ldots\}$, for the Neumann condition. Moreover, the letters (N) and (D), in tables (b) and (c), represent the missing coefficients, due to the constraints on the values of the subtraction $(n-\ell)$. The letter (N) means that the respective coefficient belongs to the Neumann series, table (c), whereas the letter (D) means that the coefficient belongs to the Dirichlet series, table (b).
\end{table}

Observing the arrangement of the respective elements in these tables, we note that the Dirichlet and Neumann series make up the Einstein series. In other words, the empty spaces in Tables \ref{Tab0}-(b) and \ref{Tab0}-(c) complement each other so that their superposition (sum) is equal to the elements in Table \ref{Tab0}-(a). Precisely, we note that
\begin{eqnarray}\label{Sec2eq04a}
	\text{W}^{\text{(D)}} + \text{W}^{\text{(N)}} = 2\text{W}^{\text{(E)}}.
\end{eqnarray}
The origin of factor 2 has already been explained previously. It is also important for the current computation to find the difference between the two solutions, that is, to calculate the quantity
\begin{eqnarray}\label{Sec2eq04b}
\text{W}^{\text{(B)}} = \text{W}^{\text{(N)}} - \text{W}^{\text{(D)}}.
\end{eqnarray}
Then, once the results of Eqs. \eqref{Sec2eq04a} and \eqref{Sec2eq04b} are obtained, we can obtain the respective expressions for the cases of Dirichlet and Neumann boundary conditions, observing that
\begin{eqnarray}\label{Sec2eq05}
\text{W}^{\text{(j)}}=\text{W}^{\text{(E)}}+\delta^{\text{(j)}}\dfrac{\text{W}^{\text{(B)}}}{2},
\end{eqnarray}
where we conveniently define $\delta^{\text{(j)}} = [\delta^{\text{(D)}},\delta^{\text{(N)}}] = [-1,+1]$. In Eqs. \eqref{Sec2eq04b} and \eqref{Sec2eq05} the subscript B refers to the contribution of the boundary condition.

According to the discussions presented above, now we need to find the quantity $\frac{\text{W}^{\text{(B)}}}{2}$, in order to obtain the respective Wightman functions $\text{W}^{\text{(D)}}$ and $\text{W}^{\text{(N)}}$ via Eq. \eqref{Sec2eq05}, since $\text{W}^{\text{(E)}}$ is given by Eqs. \eqref{eqWF_EinsteinUniverseA} and \eqref{eqWF_EinsteinUniverseB}. From the Eqs. \eqref{wightman_fucntion_FDN}, \eqref{coef_wfFDN} and \eqref{Sec2eq04b} we can write
\begin{eqnarray}\label{Sec2eq06a}
\dfrac{\text{W}^{\text{(B)}}}{2} = {\sum_{n,\ell}}^{\text{(N)}}f_{n\ell}C_{n-\ell}^{\ell+1}(\cos(\chi))C_{n-\ell}^{\ell+1}(\cos(\chi')) - {\sum_{n,\ell}}^{\text{(D)}}f_{n\ell}C_{n-\ell}^{\ell+1}(\cos(\chi))C_{n-\ell}^{\ell+1}(\cos(\chi')),
\end{eqnarray}
where we conveniently define the new coefficients
\begin{eqnarray}\label{Sec2eq06b}
f_{n\ell} = \dfrac{|\text{N}_{n\ell}^{\text{(E)}}|^{2}}{4\pi}(2\ell+1)e^{-i\omega_{n}\Delta t}\sin^{\ell}(\chi)\sin^{\ell}(\chi')P_{\ell}(\cos(\gamma)).
\end{eqnarray}
The respective sum symbols in Eq. \eqref{Sec2eq06a} are given by Eqs. \eqref{sum_symb_dirichlet} and \eqref{sum_symb_Neumann}.

Exploiting the fact that $(n-\ell)$ in the last term of Eq. \eqref{Sec2eq06a} is an odd number, we can absorb the negative sign in the Gegenbauer polynomial using the symmetry property \cite{abramowitz1970handbook} 
\begin{eqnarray}\label{sym_gegenbauer}
C_{n}^{\lambda}(-z) = (-1)^{n}C_{n}^{\lambda}(z).
\end{eqnarray}
Similarly, this procedure can be applied to the first term of Eq. \eqref{Sec2eq06a}, that is, since $(n-\ell)$ is an even number we can also introduce a negative sign in the Gegenbauer polynomial without loss of generality. Thus, based on the properties of trigonometric functions, namely, $\cos(\pi-\chi')=-\cos(\chi)$ and $\sin(\pi-\chi')=\sin(\chi)$, we can manipulate the resulting expression in oder to identify the coefficients $\text{W}^{\text{(E)}}_{n\ell}$ with $\chi'\rightarrow(\pi-\chi')$. Therefore, following the described methodology we can add both terms and obtain that
\begin{eqnarray}\label{Sec2eq07}
\dfrac{\text{W}^{\text{(B)}}}{2} &=& \sum_{n=0}^{\infty}\sum_{\ell=0}^{n}\left. \text{W}_{n\ell}^{\text{(E)}}\right|_{\chi'\rightarrow(\pi-\chi')} \nonumber \\
&=& \left. \text{W}^{\text{(E)}}\right|_{\chi'\rightarrow(\pi-\chi')}.
\end{eqnarray}

The expression above indicates that the boundary contribution $\frac{\text{W}^{\text{(B)}}}{2}$ can be obtained from the expression for the Wightman function $\text{W}^{\text{(E)}}$ referring to Einstein's universe. Therefore, according to Eqs. \eqref{Sec2eq07} e \eqref{eqWF_EinsteinUniverseA} we find that
\begin{eqnarray}\label{Boundary_ContributionA}
\dfrac{\text{W}^{\text{(B)}}}{2} = \dfrac{im_{F}}{8\pi a_{0}\sin\left(\frac{\Delta\bar{s}}{a_{0}}\right)}\sum_{n=-\infty}^{\infty}\dfrac{(\Delta\bar{s} +2\pi a_{0} n)}{\bar{\sigma}_{n}}H_{1}^{(2)}(m_{F}\bar{\sigma}_{n}),
\end{eqnarray}	
where similar to Sec.\ref{sec2C1} we have that
\begin{eqnarray}\label{Boundary_ContributionB}
\bar{\sigma}_{n}^{2} = \Delta t^{2} - (\Delta\bar{s}+2\pi a_{0} n)^{2}
\end{eqnarray}
and
\begin{eqnarray}\label{Boundary_ContributionC}
\cos(\bar{\alpha}) = \cos\left(\dfrac{\Delta\bar{s}}{a_{0}}\right)=\cos(\chi)\cos(\pi-\chi')+\sin(\chi)\sin(\pi-\chi')\cos(\gamma),
\end{eqnarray}
with $\cos(\gamma)$ given by Eq. \eqref{prop_add_theo_sphe_harm2}. Note that the massless limit of Eq. \eqref{Boundary_ContributionA} corresponds to Eq. \eqref{eqWF_EinsteinUniverseB} with the change $\Delta s\rightarrow\Delta\bar{s}$.

Similar to Ref. \cite{Ferreira:2023gzv}, here we are interested in the case of a massless scalar field. Thus, in view of Eqs. \eqref{eqWF_EinsteinUniverseB} and \eqref{Sec2eq05}, for our purposes, we can establish the following expression:
\begin{eqnarray}\label{Massless_GE_WF_EDNa}
\text{W}^{\text{(i)}}(x,x') = -\dfrac{1}{4a_{0}\pi^{2}}\sum_{n=-\infty}^{\infty}\left[ w_{n}(\Delta t,\Delta s) +\delta^{\text{(i)}}w_{n}(\Delta t,\Delta \bar{s})\right],
\end{eqnarray}
with i = (E, D, N), $\delta^{\text{(i)}}=[\delta^{\text{(E)}},\delta^{\text{(D)}},\delta^{\text{(N)}}]=[0,-1,+1],$
\begin{eqnarray}\label{Massless_GE_WF_EDNb}
w_{n}(\Delta t,\Delta s) = \dfrac{(\Delta s +2\pi a_{0}n)}{\sin\left(\frac{\Delta s}{a_{0}}\right)\sigma_{n}^{2}}
\end{eqnarray}
and
\begin{eqnarray}\label{Massless_GE_WF_EDNc}
w_{n}(\Delta t,\Delta \bar{s}) = \dfrac{(\Delta \bar{s} +2\pi a_{0}n)}{\sin\left(\frac{\Delta\bar{s}}{a_{0}}\right)\bar{\sigma}_{n}^{2}}.
\end{eqnarray}
In the above expressions, all parameters have already been defined previously. Futhermore, for general purposes, in Eq. \eqref{Massless_GE_WF_EDNa}, we include the case $\delta^{\text{(E)}}$ in order to compare and verify the consistency of the results obtained below. In fact, for the case $\delta^{\text{(E)}}=0$ we obtain the PFWF for Einstein's universe and the expressions presented below should provide the results obtained in Ref. \cite{Ferreira:2023gzv}. 

Eq. \eqref{Massless_GE_WF_EDNa} corresponds to the expression of the PFWF for a massless scalar field in Einstein's universe and in the Einstein's universe under Dirichlet and Neumann boundary conditions. 
This is a very useful expression, because we can use it to calculate the physical observables that depend on it and in the end study the particular cases by choosing the appropriate values of $\delta^{\text{(i)}}$. In fact, it is important to note that the operations only affect the $w_{n}$ functions, as the $\delta^{\text{(i)}}$ coefficients are constant.
A similar structure can be found in Ref. \cite{Ferreira:2023uxs} for a massless scalar field in the presence of two parallel planes in Minkowski spacetime. 
In the next sections we will use Eq. \eqref{Massless_GE_WF_EDNa} to study the momentum dispersion of a point particle induced by the quantum vacuum fluctuations of a massless scalar field in the case of the Einstein's universe subject to Dirichlet and Neumann boundary conditions.

\section{Particle dynamics and momentum dispersion}\label{sec3}

\subsection{General expresions}\label{sec3.1}

%
In curved spacetime, the four momentum $p^{\mu}=m_{\text{p}}u^{\mu}$ of a point particle of mass $m_{\text{p}}$ coupled to a massless scalar field $\psi(x)$, is governed by the equation of motion \cite{poisson2011motion,Ferreira:2023gzv,adler2021general}
\begin{equation}\label{Sec3eq00}
\dfrac{dp^{\mu}}{d\tau} + m_{\text{p}}\Gamma_{\nu\rho}^{\mu}u^{\nu}u^{\rho}=-qg^{\mu\nu}\nabla_{\nu}\psi(x),
\end{equation}
%
%
where the constant $q$ corresponds to the charge of the particle or, in other words, the strength of coupling between the particle and the scalar field. In Eq. \eqref{Sec3eq00}, $u^{\mu} = dx^{\mu}/d\tau$ represents the velocity four-vector of the particle, defined in the usual way  as the rate of variation of the spacetime coordinates $x^{\mu}$ with respect to proper time $\tau$. The $\Gamma_{\nu\rho}^{\mu}$ symbols are duly defined in Eq. \eqref{ChristoffelSymbols}. In addition, the mass of the particle in the above equations corresponds to a dynamic mass, which can vary with time due to it is interaction with the scalar field, and is defined by the relations \cite{poisson2011motion}
\begin{eqnarray}\label{Sec3eq01a}
m_{\text{p}}(\tau)=m_{0} -q\psi(x)
\end{eqnarray}
and
\begin{eqnarray}\label{Sec3eq01b}
\dfrac{dm_{\text{p}}}{d\tau}=-qu^{\mu}\nabla_{\mu}\psi(x),
\end{eqnarray}
%
%
where $m_{0}$ represents the bare mass of the particle, that is, in the absence of interaction effects with the scalar field. In fact, it is noted that in the limit $q\rightarrow0$ we obtain that $m_{\text{p}}=m_{0}$.

%
For simplicity, we perform the present study considering a non-relativistic limit, which allows us to ignore the possibles effects of the particle backreaction. This scenario is compatible with the point particle regime that we are considering \cite{mota2020induced}. Then, in this validity regime, the proper and coordinate time are approximately equal and also only the spatial components of Eq. \eqref{Sec3eq00} become significant, allowing us to establish the following expression: 
\begin{eqnarray}\label{Sec3eq02}
\dfrac{dp^{i}}{dt} + m_{\text{p}}\Gamma_{\nu\rho}^{i}u^{\nu}u^{\rho}=-qg^{i\nu}\nabla_{\nu}\psi(x) + f_{\textrm{ext}}^{i}.
\end{eqnarray}
%
%
The term $f_{\textrm{ext}}^{i}$ has been added in order to include the effect of sources external to the system (composed here of the particle and the scalar field). It is instructive to observe that this expression corresponds to a generalization for the Newton's second law to an arbitrary coordinate system \cite{adler2021general}. In addition, this is a classical equation.


In geral, quantum fields permeating curved spacetimes, that is, in the presence of gravitational effects, can naturally suffer backreaction effects. In fact, the nontrivial curvature of spacetime (gravitity), even classically, modifies the vacuum fluctuations of the fields. In turn, this modification gives rise to a renormalized vacuum energy, which is provide by VEV of the renormalized energy-momentum tensor (EMT) \cite{mukhanov2007introduction}. Consequently, according to Einstein's field equations, this nonzero VEV to the EMT can contribute as a source of gravity (or ``spacetime curvature''), which can be encoded into the metric of the concerned spacetime. An example which illustrates this process can be seen in Ref. \cite{DeLorenci:2008nr} where, among other things, the authors obtain quantum corrections for the metric tensor of a spinning cosmic string in the presence of a conformally coupled massless scalar field. Furthermore, regarding backreaction effects in Einstein's universe, these were discussed in Refs. \cite{Altaie:2001vv} and \cite{Altaie:2002tv} from a different perspective than the one we reported in this paragraph. In view of the above, we should point out that, in our expressions, backreaction effects directly influence Eq. \eqref{Sec3eq02} through the quantities $g^{ij}$ and $\Gamma_{\nu\rho}^{i}$. However, in the present case, such effects are neglected, due to the nonrelativistic regime and the assumed point particle treatment.

%
Despite the simplifications introduced by the hypothesis discussed above, we perceives that solving Eq. \eqref{Sec3eq02} is still a difficult task. The reason is that the non-zero Christoffel symbols $\Gamma_{\nu\rho}^{i}$ for the metric \eqref{metricEinsteinUniverse} (shown in Table \ref{tableCSEuniverse}) produces a constraint between the distinct components of $p^{i}$ and $u^{\mu}$, making the direct solution of such expressions difficult. However, based on the fact that the origin of $\Gamma_{\nu\rho}^{i}$  is geometric, we can interpret the second term on the right hand side of Eq. \eqref{Sec3eq02} as a classical force. Therefore, in order to analyze only the contributions arising from quantum field fluctuations to the motion of particle, it is appropriate to suppose that 
\begin{eqnarray}\label{Sec3eq03}
f_{\textrm{ext}}^{i} = m_{\text{p}}\Gamma_{\nu\rho}^{i}u^{\nu}u^{\rho}.
\end{eqnarray}
%
%
Another possible justification for the expression of force $f_{\textrm{ext}}^{i}$ chosen above is that, in the present approach,  we are considering that the quantum effects are exclusively from the massless scalar field $\psi(x)$. 
%
%
In order to better understand the meaning of $f_{\text{ext}}^{i}$, as well as the level of approximation we are considering, it is opportune to first observe how geometry affects Eq. \eqref{Sec3eq02} and consequently the IQBM of the particle. 

%
The dynamics of the point particle, given by Eq. \eqref{Sec3eq02}, is affected by the geometric coefficients $g^{ij}$ and $\Gamma_{\nu\rho}^{i}$, which depend on the coordinate system used to describe the spacetime geometry. In addition, it is also affected by the modes of the scalar field $\psi$, which propagates throughout the entire expanse of spacetime, probing information about the geometric structure of Einstein's universe. This is easily seen by observing, for instance, at the nontrivial mathematical structures that the expressions for the modes \eqref{Sec2eq00a} and \eqref{modesD&N} have, which differ from the usual form of plane waves $e^{-i\omega t+i\mathbf{k}\cdot\mathbf{x}}$ in Minkowski spacetime. 

%
Certainly, solution of Eq. \eqref{Sec3eq02} considering $f_{\textrm{ext}}^{i}=0$ would give us a more complete description for the $p^{i}$ components of the momentum, but this case becomes unfeasible to be solved analytically. Then, for mathematical reasons, Eq. \eqref{Sec3eq03} is crucial, as its implementation allows us to find analytical solutions. From a physical viewpoint, considering Eq. \eqref{Sec3eq03} in our calculations means that we are assuming a regime in which the geometric effects of spacetime locally do not affect the particle dynamics. In this sense, we are assuming that the particle indirectly probes the nontrivial effect of the geometry of Einstein's universe through the quantum fluctuations of the scalar field, which permeates the entire spacetime.
\begin{table}[h]
\caption{Non-zero Christoffel symbols $\Gamma_{\nu\rho}^{\mu}$ for Einstein's universe.}\label{tableCSEuniverse}
	\begin{tabular}{ccc}
	\toprule\toprule
	 	$\Gamma^{\chi}_{\theta\theta}$ &\hspace{2cm} & $-\sin(\chi)\cos(\chi)$ \\ \midrule
	  	$\Gamma^{\chi}_{\phi\phi}$ & & $-\cos(\chi)\sin(\chi)\sin^{2}(\theta)$\hspace{0.5cm} \\ \midrule
	  	\hspace{0.5cm}$\Gamma^{\theta}_{\chi\theta}, \Gamma^{\theta}_{\theta\chi},\Gamma^{\phi}_{\chi\phi},\Gamma^{\phi}_{\phi\chi}$ & & $\cot(\chi)$ \\ \midrule
	  	$\Gamma^{\theta}_{\phi\phi}$ & & $-\sin(\theta)\cos(\theta)$ \\ \midrule
	  	$\Gamma^{\phi}_{\theta\phi},	\Gamma^{\phi}_{\phi\theta}$ & & $\cot(\theta)$\\
	\bottomrule\bottomrule
	\end{tabular}
\end{table}
%
%
%

%
In view of Eqs. \eqref{Sec3eq02} and \eqref{Sec3eq03}, it follows that momentum $p^{i}$ obeys the first order differential equation 
\begin{eqnarray}\nonumber
\dfrac{dp^{i}}{dt}=-qg^{i\nu}\nabla_{\nu}\psi(x),
\end{eqnarray}
%
%
whose integration between two successive instants of time, $t_{0}=0$ and $t=\tau$, subject to classical condition of initial value $p^{i}(0)=0$, gives as the expression
\begin{eqnarray}\label{Sec3eq04}
p^{i}(x)=-q\int_{0}^{\tau}dt g^{i\nu}\nabla_{\nu}\psi(x).
\end{eqnarray}
%
%
This equation provides the non-relativistic expressions that govern the behavior of the spatial components of the momentum of the point particle of charge $q$, coupled to a massless scalar field $\psi(x)$ in an arbitrary spacetime, described by the set of coordinate $x$. We emphasize that in the present case $x=(t,\chi,\theta,\phi)$. Furthermore, we will consider that the temporal variation of the coordinates is small enough that we can neglect them.

%
The dispersion or variance of an observable $\hat{\mathcal{O}}$, with respect to an arbitrary quantum state $|\Omega \rangle$, is defined by the relation \cite{sakurai1994} 
\begin{eqnarray}\label{Sec3eq05}
\langle(\Delta\hat{\mathcal{O}})^{2}\rangle=\langle\hat{\mathcal{O}}^{2}\rangle-\langle\hat{\mathcal{O}}\rangle^{2},
\end{eqnarray}
where we use the compact notation $\langle\ldots\rangle=\langle\Omega|\ldots|\Omega\rangle$ for the expected values.
The first term on the right hand side is the expectation value (or mean value) of squared $\hat{\mathcal{O}}$ operator and the second one is the squared of the expectation value of $\hat{\mathcal{O}}$. The relation in \eqref{Sec3eq05} is general and holds for all cases, that is, for any quantum state $|\Omega\rangle$ and operator $\hat{\mathcal{O}}$, but here we are interested in the state of quantum vacuum so that $|\Omega\rangle = |0\rangle$. 
So, quantizing Eq. \eqref{Sec3eq04}, through the quantum prescription $(p^{i},\psi)\rightarrow(\hat{p}^{i},\hat{\psi})$ and observing that $\langle\hat{p}^{i}\rangle=0$, as a result of Eqs. \eqref{GEFieldOperator} and \eqref{Sec2VacuumAlgebra}, from Eqs. \eqref{Sec3eq04} and \eqref{Sec3eq05} it follows that
\begin{eqnarray}\label{Sec3eq06}
\langle(\Delta \hat{p}^{i})^{2}\rangle_{\textrm{ren}}^{\textrm{(j)}} = \lim_{x'\rightarrow x}q^{2}\int_{0}^{\tau}dt'\int_{0}^{\tau}dt g^{ii}(x)g^{ii}(x')\dfrac{\partial^{2}\textrm{W}_\textrm{ren}^{\textrm{(j)}}(x,x')}{\partial x^{i}\partial x'^{i}},
\end{eqnarray}
where we identify the PFWF $\langle 0|\hat{\psi}(x)\hat{\psi}(x') |0\rangle=\textrm{W}(x,x')$, according to definition \eqref{GEwightmanfunction}. It is noted that, since $\langle(\Delta \hat{p}^{i})^{2}\rangle_{\textrm{ren}}^{\text{(j)}}=\langle(\hat{p}^{i})^{2}\rangle_{\textrm{ren}}^{\text{(j)}}$, here the concepts of dispersion and VEV of the squared momentum are synonymous. In addition, to establish the corresponding expression of $\langle(\Delta \hat{p}^{i})^{2}\rangle_{\textrm{ren}}^{\textrm{(j)}}$ we use the fact that the metric tensor of Einstein's universe,  Eq. \eqref{metricEinsteinUniverse}, is diagonal.

%
Eq. \eqref{Sec3eq06} will allow us to calculate the dispersion of the particle momentum components induced by the quantum vacuum fluctuations of the massless scalar field in Einstein's universe, in the presence or absence of boundary conditions. For purposes of clarity, before we proceed, it is instructive to make some comments about the structure of this expression. The formal limit notation represents the well-known operation of the coincidence limit, which must be performed in our calculations in order to obtain the desired quantity. For simplicity, from now on, this operation will be omitted, leaving it implicit. The indices $i$ and j, respectively, define the spatial coordinate and boundary condition considered in the analysis, that is, $i=(\chi,\theta,\phi)$ and j = (E, D, N). The subscript ``ren'' indicates that the quantities used must be duly renormalized. In the present case, as indicated in Refs. \cite{dowker1976covariant,dowker1977vacuum, Ozcan:2006jn}, the regularization of the observables is performed by discarding the term $w_{0}(\Delta t,\Delta s)$ from Eq. \eqref{Massless_GE_WF_EDNa}, which is the divergent contribution in the coincidence limit $(\Delta t,\Delta s)\rightarrow 0$.

\subsection{Dispersion for the momentum components}\label{sec3.2}

 We will now use the results from Sections \ref{sec2C2} and \ref{sec3.1} to study the behavior of the momentum dispersion components, or equivalently the VEV of the components of momentum squared. Essentially, this will be accomplished by using Eqs. \eqref{Massless_GE_WF_EDNa} and \eqref{Sec3eq06}. In fact, by making use of Eq. \eqref{Sec3eq06}, for each selected component $i=(\chi,\theta,\phi)$, after identifying the metric elements $g^{ii}$, we also use Eq. \eqref{Massless_GE_WF_EDNa} and calculate the integrals and derivatives in Eq. \eqref{Sec3eq06}. Then, we specify the constant coefficient $\delta^{\text{(j)}}$ to observe the momentum dispersion component referring to the respective boundary condition case. The metric coefficients $g^{ii}$ can be easily and directly obtained from Eq. \eqref{metricEinsteinUniverse} using the identity $g_{\mu\nu}g^{\mu\nu}=1$, since there are no cross terms in the line element, so that $g^{ii}=\left\{g^{\chi\chi}; g^{\theta\theta}; g^{\phi\phi}\right\} = -a_{0}^{-2}\left\{1; \sin^{-2}\chi; \sin^{-2}\chi\sin^{-2}\theta\right\}$. Here, we faithfully follow the notation and mathematical structures introduced in Ref. \cite{Ferreira:2023gzv}.

First, considering the component $i=\chi$, according to the discussions above, we have that 
\begin{eqnarray}\label{Sec3eq07}
\langle(\Delta \hat{p}^{\chi})^{2}\rangle_{\textrm{ren}}^{\textrm{(j)}} = 2q^{2}a_{0}^{-4}\int_{0}^{\tau}d\eta(\tau-\eta) \left[K_{\chi}^{\textrm{(E)}}(x,x')+\delta^{\textrm{(j)}}K_{\chi}^{\textrm{(B)}}(x,x') \right],
\end{eqnarray}
where to reduce the double integral in Eq. \eqref{Sec3eq06} we use the  identity \cite{de2014quantum,camargo2018vacuum}
\begin{eqnarray}\label{Sec3eq_integral_identity}
\int_{0}^{\tau}dt'\int_{0}^{\tau}dt\mathcal{G}(|t-t'|) = 2\int_{0}^{\tau}d\eta(\tau-\eta)\mathcal{G}(\eta), 
\end{eqnarray}
such that $\eta=|t-t'|$, and to simplify Eq. \eqref{Sec3eq07} we define the integral kernels
\begin{eqnarray}\label{Sec3eq_kernel_E}
K_{i}^{\text{(E)}}(x,x'):=\left.\left[\partial_{i}\partial_{i'}\text{W}_{\textrm{ren}}^{\text{(E)}}(x,x')\right]\right|_{x'=x}
\end{eqnarray}	
and 
\begin{eqnarray}\label{Sec3eq_kernel_B}
K_{i}^{\text{(B)}}(x,x'):=\left.\left[\partial_{i}\partial_{i'}\frac{\text{W}^{\text{(B)}}(x,x')}{2}\right]\right|_{x'=x}.
\end{eqnarray}
Eq. \eqref{Sec3eq_kernel_E} shows that the renormalization information is concentrated in Einstein's universe kernel, which corresponds to the first term on the right-hand side of Eq. \eqref{Massless_GE_WF_EDNa}, or Eq. \eqref{eqWF_EinsteinUniverseB}, whereas the contribution of the boundary condition kernel, Eq. \eqref{Sec2eq07}, is given by second term of the same expression. From Eqs. \eqref{Sec3eq07}, \eqref{Sec3eq_kernel_E} and \eqref{Sec3eq_kernel_B}, we note that referring to each kernel $K_{i}^{(k)}(x,x')$ we have an integral
\begin{eqnarray}\label{Sec3eq_integral_kernel}
I_{i}^{(k)}(x,x'):=\int_{0}^{\tau}d\eta(\tau-\eta)K_{i}^{(k)}(x,x'),
\end{eqnarray}	
with $k =$ (E, B).

As we observe from Eq. \eqref{Sec3eq07}, the momentum dispersion $\langle(\Delta \hat{p}^{\chi})^{2}\rangle_{\textrm{ren}}^{\textrm{(j)}}$ is composed of the sum of two contributions. This aspect is shared by all components of the momentum dispersion and is a direct consequence of the mathematical structure of the PFWF. In fact, observing Eqs. \eqref{Massless_GE_WF_EDNa} and \eqref{Sec3eq06}  we note that the dispersion of the momentum components allows the decomposition
\begin{eqnarray}\label{Sec3eq08}
\langle(\Delta \hat{p}^{i})^{2}\rangle_{\textrm{ren}}^{\text{(j)}} = \langle(\Delta \hat{p}^{i})^{2}\rangle_{\textrm{ren}}^{\text{(E)}}+\delta^{\text{(j)}}\langle(\Delta \hat{p}^{i})^{2}\rangle^{\text{(B)}},
\end{eqnarray}
which shows that, for the case of the Einstein's universe with boundary conditions, the momentum dispersion is composed by two contributions, one coming from Einstein's universe and the other from the boundary conditions used. From Eq. \eqref{Sec3eq08} we can immediately see that the present study generalizes and complements the previous investigations \cite{Ferreira:2023gzv}, in addition to it showing the distinctions between the two studies. In addition, this relation show us that we can calculate the two contributions to dispersion separately.

From Eqs. \eqref{Sec3eq07}, \eqref{Sec3eq_kernel_E}, \eqref{Sec3eq_kernel_B} and \eqref{Sec3eq_integral_kernel}, after some simplifications, we find that the dispersion of the physical momentum related to the $\chi$ component in the coincidence limit is given by
\begin{eqnarray}\label{Sec3eq_physical_momentum_disp_chi}
\langle(\Delta \hat{\mathcal{p}}^{\chi})^{2}\rangle_{\textrm{ren}}^{\textrm{(j)}} = \dfrac{2q^{2}}{a_{0}^{2}}\left[I_{\chi}^{\textrm{(E)}}+\delta^{\textrm{(j)}}I_{\chi}^{\textrm{(B)}}\right],
\end{eqnarray}
with
\begin{eqnarray}\label{Sec3eq_physical_momentum_disp_chi_E}
I_{\chi}^{\textrm{(E)}} = -\dfrac{1}{(12\pi)^{2}}\left\{1+\frac{12}{\tau_{a}^{2}}-3\csc^{2}\left(\dfrac{\tau_{a}}{2}\right) +6\ln\left[\dfrac{\sin\left(\frac{\tau_{a}}{2}\right)}{\left(\frac{\tau_{a}}{2}\right)}\right]^{2}  \right\}
\end{eqnarray}	
and
\begin{eqnarray}\label{Sec3eq_physical_momentum_disp_chi_B}
I_{\chi}^{\textrm{(B)}} = \dfrac{\sec^{2}(\chi)\sin^{2}\left(\frac{\tau_{a}}{2}\right)}{8\pi^{2}\left[\cos(\tau_{a})+\cos(2\chi)\right]} - \dfrac{\cot(2\chi)\csc(2\chi)}{8\pi^{2}}\ln\left[\dfrac{2\cos^{2}(\chi)}{\cos(\tau_{a})+\cos(2\chi)}\right]^{2} + \dfrac{[3+\cos(4\chi)]\csc^{3}(2\chi)}{8\pi^{2}}\Delta \mathcal{P},
\end{eqnarray}
where for practicality we define the auxiliary functions
\begin{eqnarray}\label{Sec3eq_physical_momentum_disp_chi_aux_funA}
\Delta \mathcal{P} = u^{+}_{f}\ln\left|\dfrac{\sec(u^{+}_{f})}{\sec(u^{+}_{i})}\right| - u^{-}_{f}\ln\left|\dfrac{\sec(u^{-}_{f})}{\sec(u^{-}_{i})}\right| - \Delta\mathcal{F}_{f} + \Delta\mathcal{F}_{i},
\end{eqnarray}
\begin{eqnarray}\label{Sec3eq_physical_momentum_disp_chi_aux_funB}
\Delta\mathcal{F}_{f}= \mathcal{F}(u^{+}_{f}) -\mathcal{F}(u^{-}_{f}),
\end{eqnarray}
\begin{eqnarray}\label{Sec3eq_physical_momentum_disp_chi_aux_funC}
\Delta\mathcal{F}_{i}= \mathcal{F}(u^{+}_{i}) -\mathcal{F}(u^{-}_{i})
\end{eqnarray}
and
\begin{eqnarray}\label{Sec3eq_physical_momentum_disp_chi_aux_funD}
\mathcal{F}(\xi)= \dfrac{1}{2}\ln\left(\dfrac{|\sec(\xi)|}{2}\right)\arctan[\tan(\xi)] -\dfrac{i}{4}\left\{\textrm{Li}_{2}\left[\dfrac{1+i\tan(\xi)}{2}\right]-\textrm{Li}_{2}\left[\dfrac{1-i\tan(\xi)}{2}\right] \right\},
\end{eqnarray}
with $u^{\pm}_{f} = \tau_{a}/2\pm\chi$, $u^{\pm}_{i}=\pm\chi$ and $\tau_{a}=\tau/a_{0}$ the dimensionless time. In Eq. \eqref{Sec3eq_physical_momentum_disp_chi_aux_funD} $\textrm{Li}_{\nu}(z)$ corresponds to the polylogarithmic function in the variable $z$, such that $|z|<1$, which is related to the Lerch function $\Phi(z,\nu,v)$ by the relation $\textrm{Li}_{\nu}(z)=z\Phi(z,\nu,1)$, where $v\neq 0,-1,-2,\ldots $ \cite{weisstein}.

\begin{figure}[h]
\centering
\includegraphics[scale=0.5]{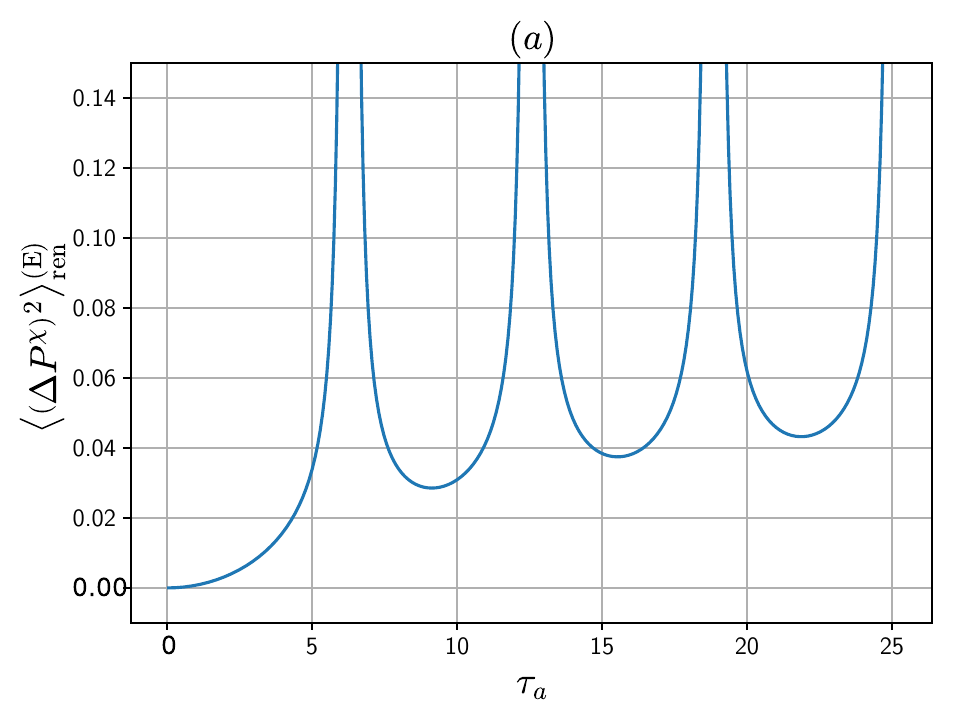}
\includegraphics[scale=0.5]{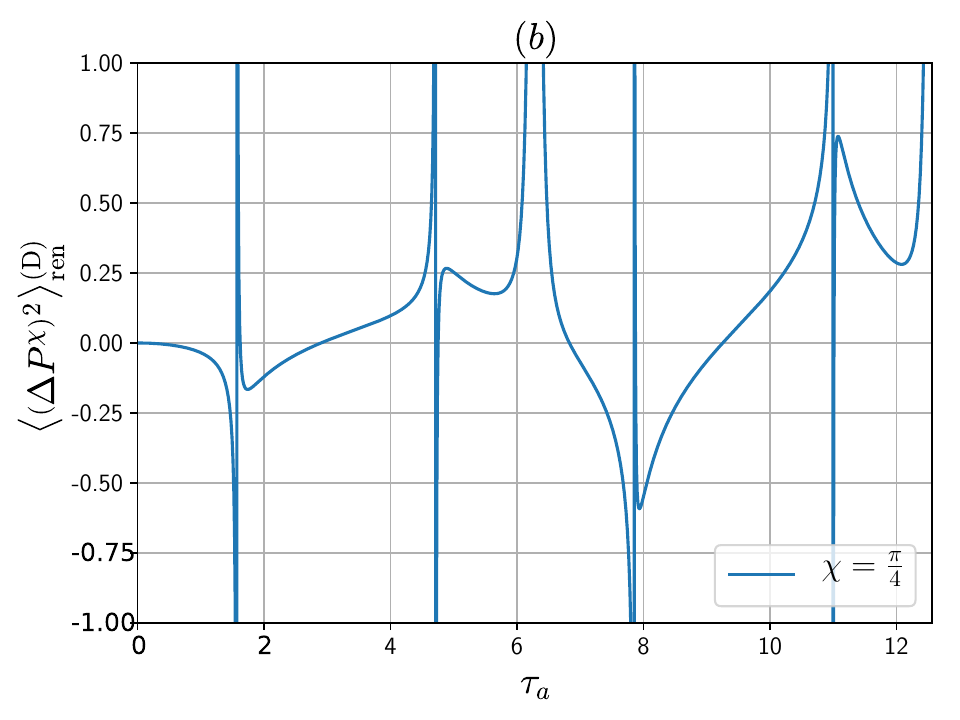}\includegraphics[scale=0.5]{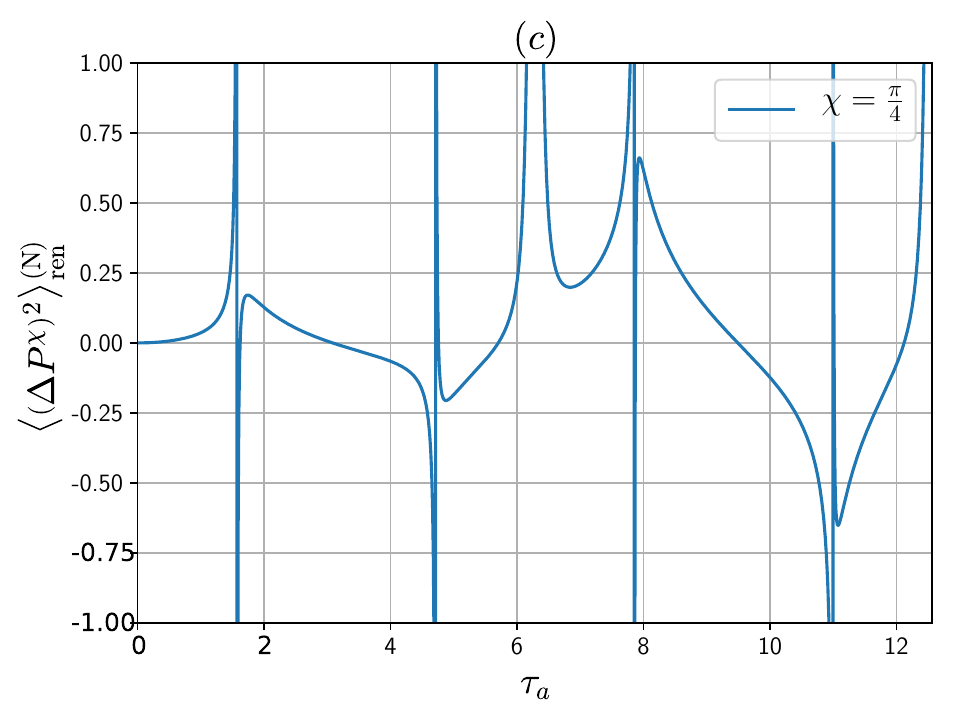}
\caption{Behavior of the renormalized dimensionless dispersion of the $\chi$ component of the physical momentum as a function of the dimensionless time $\tau_{a}$, for a point particle coupled to a real massless scalar field, in the cases of (a) Einstein's universe and the Einstein's universe with (b) Dirichlet and (c) Neumann boundary conditions. Note that we have defined the dimensionless quantity $\langle(\Delta P^{\chi})^{2}\rangle_{\textrm{ren}}^{\textrm{(j)}}=\left(\frac{a_{0}}{q}\right)^{2}\langle(\Delta \hat{\mathcal{p}}^{\chi})^{2}\rangle_{\textrm{ren}}^{\textrm{(j)}}$. Figures (b) and (c) assume the particular value $\chi=\frac{\pi}{4}$.}\label{fig01_mvsm_chi}
\end{figure}

Before we proceed let us briefly discuss some points about the above results. First, it is important to emphasize that to establish Eq. \eqref{Sec3eq_physical_momentum_disp_chi} we have used the relation between the components of the physical mometum $\hat{\mathcal{p}}^{i}$ and the coordinate momentum $\hat{p}^{i}$, which are shown in Table \ref{tab_phy_and_coord_momentum}. These relations can be deduced direcly from the line element \eqref{metricEinsteinUniverse}. The graphical behavior of Eq. \eqref{Sec3eq_physical_momentum_disp_chi} as a function of $\tau_{a}$ is shown in Figure \ref{fig01_mvsm_chi}.

To obtain the contribution $I_{\chi}^{\textrm{(E)}}$ initially we perform the sum and the coincidence limit $(\theta',\phi')\rightarrow(\theta,\phi)$ in advance. Then, as indicated by Eq. \eqref{Sec3eq_kernel_E}, we successively derive with respect to the pair of variables $(\chi,\chi')$ and take $\chi'=\chi$, so that in the end we obtain the kernel $K_{\chi}^{\text{(E)}}$. Finaly, we perform the integral $I_{\chi}^{\textrm{(E)}}$, defined by Eq. \eqref{Sec3eq_integral_kernel}, and find the result shown in Eq. \eqref{Sec3eq_physical_momentum_disp_chi_E}.
For the boundary term $I_{\chi}^{\textrm{(B)}}$ we follow a similar routine, but in a different order. Initially we perform the coincidence limit on the variables $\theta$ and $\phi$, then we derive with respect to $(\chi,\chi')$, take $\chi=\chi'$ and simplify the resulting expression considering the condition $0\leq\chi\leq\frac{\pi}{2}$. We must remember that for the bounded case $\chi=\left[0,\frac{\pi}{2}\right]$. Lastly, we compute the sums and integrals for each term, obtaining as a result Eq. \eqref{Sec3eq_physical_momentum_disp_chi_B}. It is important to point out that a mathematical software was also used to simplify the expressions.
\begin{table}[h]
\centering
\caption{Relation between the components of the physical momenta and the coordinate momenta.}\label{tab_phy_and_coord_momentum}
	\begin{tabular}{ccc}
	\toprule
		\hspace{0.5cm} physical momentum $\mathcal{p}^{i}$ & \hspace{2cm} & coordinate momentum $p^{i}$\hspace{0.5cm} \\ \midrule\midrule
	 	$\mathcal{p}^{\chi}$	&& 	$a_{0}p^{\chi}$ 	  	\\ \midrule
	  	$\mathcal{p}^{\theta}$ 	&& $a_{0}\sin(\chi)p^{\theta}$\\ \midrule
	  	$\mathcal{p}^{\phi}$ 	&& $a_{0}\sin(\chi)\sin(\theta)p^{\phi}$ \\ 
	\bottomrule\bottomrule
	\end{tabular}
\end{table}

\begin{figure}[h]
\centering
\includegraphics[scale=0.5]{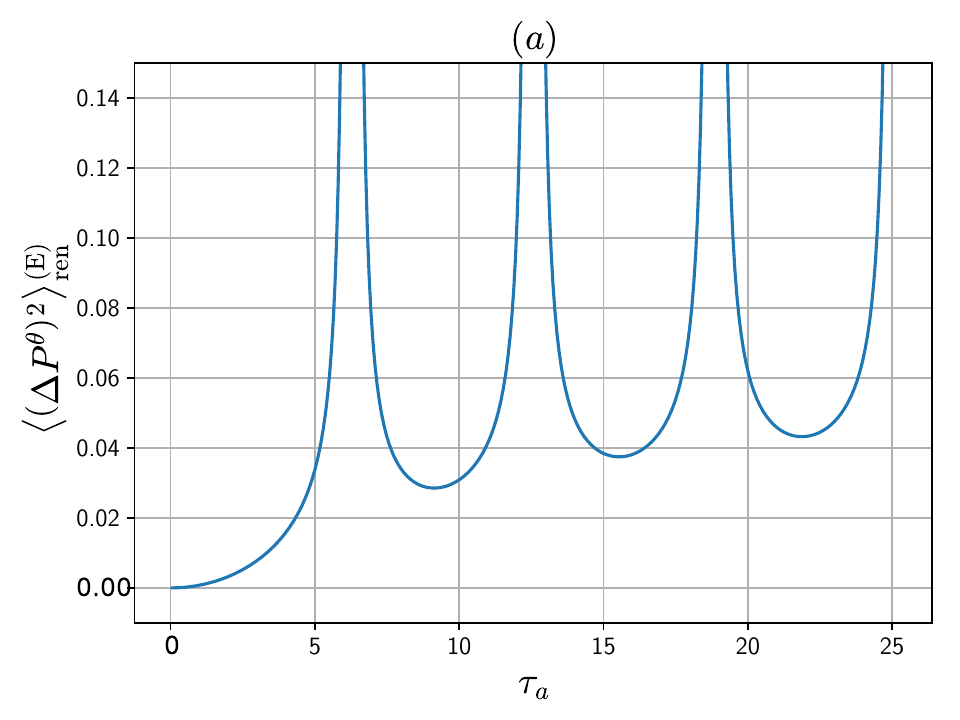}
\includegraphics[scale=0.5]{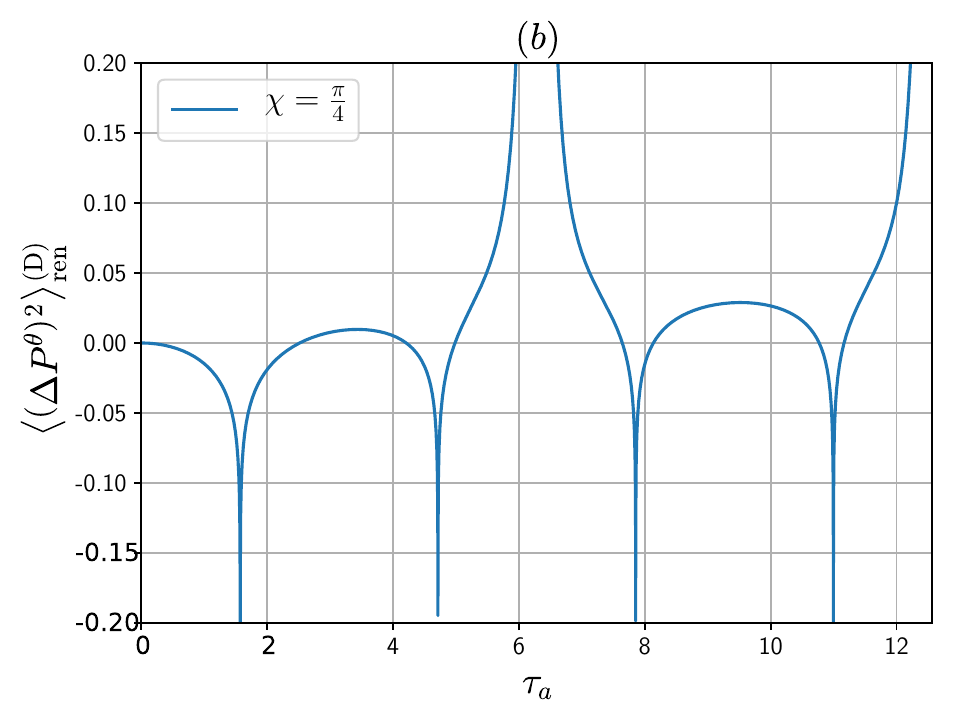}\includegraphics[scale=0.5]{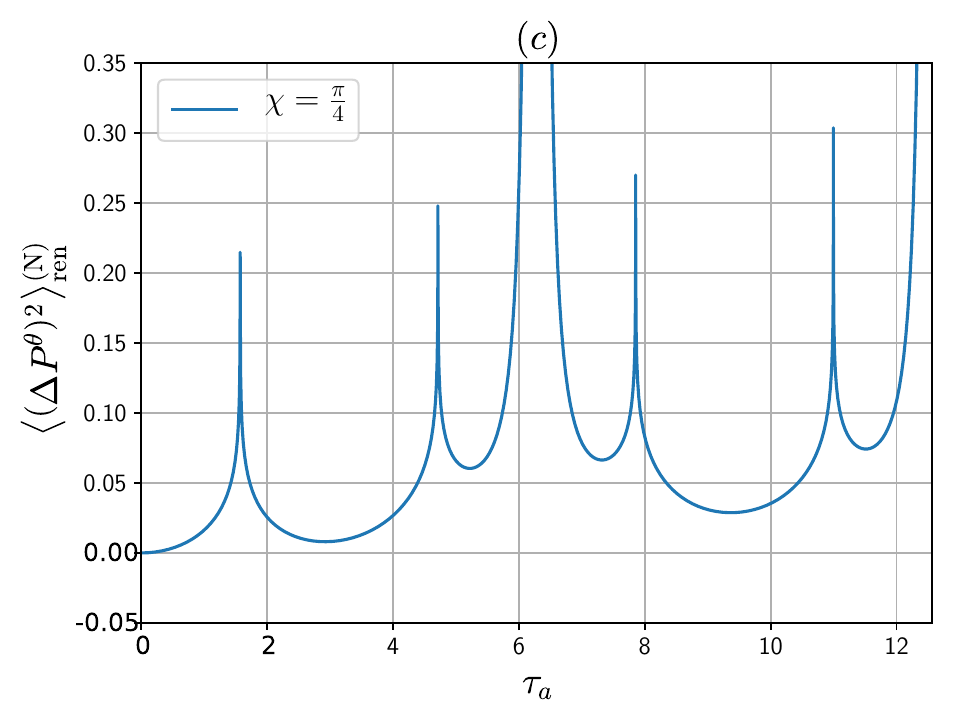}
\caption{Behavior of the renormalized dimensionless dispersion of the $\theta$ component of the physical momentum as a function of the dimensionless time $\tau_{a}$, for a point particle coupled to a real massless scalar field, in the cases of (a) Einstein's universe and the Einstein's universe with (b) Dirichlet and (c) Neumann boundary conditions. The graphics are in units of $\langle(\Delta P^{\theta})^{2}\rangle_{\textrm{ren}}^{\textrm{(j)}}=\left(\frac{a_{0}}{q}\right)^{2}\langle(\Delta \hat{\mathcal{p}}^{\theta})^{2}\rangle_{\textrm{ren}}^{\textrm{(j)}}$. Figures (b) and (c) assume the particular value $\chi=\frac{\pi}{4}$. The peaks represent divergent points. The $\phi$ component of the physical momentum dispersion, $\langle(\Delta P^{\phi})^{2}\rangle_{\textrm{ren}}^{\textrm{(j)}}$, has a similar behavior.}\label{fig02_mvsm_theta_and_phi}
\end{figure}

On the other hand, choosing $i=\theta$ in Eq. \eqref{Sec3eq06} and using Eq. \eqref{Sec3eq_integral_identity} and the definitions \eqref{Sec3eq_kernel_E} and \eqref{Sec3eq_kernel_B}, we find that the dispersion for the corresponding momentum coordinate is given by the relation
\begin{eqnarray}\label{Sec3eq09}
\langle(\Delta \hat{p}^{\theta})^{2}\rangle_{\textrm{ren}}^{\textrm{(j)}} = 2q^{2}a_{0}^{-4}\sin^{-4}(\chi)\int_{0}^{\tau}d\eta(\tau-\eta) \left[K_{\theta}^{\textrm{(E)}}(x,x')+\delta^{\textrm{(j)}}K_{\theta}^{\textrm{(B)}}(x,x') \right].
\end{eqnarray}
The solution to the above equation provide us the following expression for the dispersion of physical momentum related to the theta coordinate in the coincidence limit:
\begin{eqnarray}\label{Sec3eq_physical_momentum_disp_theta}
\langle(\Delta \mathcal{\hat{p}}^{\theta})^{2}\rangle_{\textrm{ren}}^{\textrm{(j)}} = \dfrac{2q^{2}}{a_{0}^{2}\sin^{2}(\chi)}\left[ I_{\theta}^{\textrm{(E)}}+\delta^{\textrm{(j)}}I_{\theta}^{\textrm{(B)}}\right],
\end{eqnarray}
with
\begin{eqnarray}\label{Sec3eq_physical_momentum_disp_theta_E}
I_{\theta}^{\textrm{(E)}}=\sin^{2}(\chi)I_{\chi}^{\textrm{(E)}}
\end{eqnarray}
and
\begin{eqnarray}\label{Sec3eq_physical_momentum_disp_theta_B}
I_{\theta}^{\textrm{(B)}}=\dfrac{\sec^{2}(\chi)}{(8\pi)^{2}}\ln\left[\dfrac{2\cos^{2}(\chi)}{\cos(\tau_{a})+\cos(2\chi)}\right]^{2} - \dfrac{\sec^{2}(\chi)\cot(2\chi)}{(4\pi)^{2}}\Delta\mathcal{P}.
\end{eqnarray}
The methodology used in the solutions of the integral contributions $I_{\theta}^{\textrm{(E)}}$ and $I_{\theta}^{\textrm{(B)}}$  is similar to that described for the $i=\chi$ component. The functions $I_{\chi}^{\textrm{(E)}}$ and $\Delta\mathcal{P}$ in Eqs. \eqref{Sec3eq_physical_momentum_disp_theta_E} and \eqref{Sec3eq_physical_momentum_disp_theta_B} are given by Eqs. \eqref{Sec3eq_physical_momentum_disp_chi_E} and \eqref{Sec3eq_physical_momentum_disp_chi_aux_funA}, respectively. In addition, we again use the relation between the quantities $p^{i}$ and $\mathcal{p}^{i}$, shown in Table \ref{tab_phy_and_coord_momentum}, in order to establish the dispersion of the physical momentum $\mathcal{p}^{\theta}$. The behavior of $\langle(\Delta \hat{\mathcal{p}}^{\theta})^{2}\rangle_{\textrm{ren}}^{\textrm{(j)}} $ as a function of $\tau_{a}$ is shown in Figure \ref{fig02_mvsm_theta_and_phi}.

Finally, analogously to previous cases, taking $i=\phi$ in Eq. \eqref{Sec3eq06}, using Eq. \eqref{Sec3eq_integral_identity} and identifying the kernels \eqref{Sec3eq_kernel_E} and \eqref{Sec3eq_kernel_B} we obtain that
\begin{eqnarray}\label{Sec3eq10}
\langle(\Delta \hat{p}^{\phi})^{2}\rangle_{\textrm{ren}}^{\textrm{(j)}} = 2q^{2}a_{0}^{-4}\sin^{-4}(\chi)\sin^{-4}(\theta)\int_{0}^{\tau}d\eta(\tau-\eta) \left[K_{\phi}^{\textrm{(E)}}(x,x')+\delta^{\textrm{(j)}}K_{\phi}^{\textrm{(B)}}(x,x') \right].
\end{eqnarray}
Now, calculating the respective integral contributions and using the relations in Table \ref{tab_phy_and_coord_momentum}, we find for the dispersion of the physical momentum referring to the $\phi$ component in the coincidence limit
\begin{eqnarray}\label{Sec3eq_physical_momentum_disp_phi}
 \langle(\Delta \mathcal{\hat{p}}^{\phi})^{2}\rangle_{\textrm{ren}}^{\textrm{(j)}}  = \langle(\Delta \mathcal{\hat{p}}^{\theta})^{2}\rangle_{\textrm{ren}}^{\textrm{(j)}}.
\end{eqnarray}
This result is a direct consequence of the relations between the integrals $I_{i}^{(k)}$ for the components $\theta$ and $\phi$, namely,
\begin{eqnarray}\label{Sec3eq_physical_momentum_disp_phi_E}
I_{\phi}^{\textrm{(E)}}=\sin^{2}(\theta)I_{\theta}^{\textrm{(E)}}
\end{eqnarray}
and
\begin{eqnarray}\label{Sec3eq_physical_momentum_disp_phi_B}
I_{\phi}^{\textrm{(B)}}=\sin^{2}(\theta)I_{\theta}^{\textrm{(B)}}.
\end{eqnarray}
Obviously, in view of Eq. \eqref{Sec3eq_physical_momentum_disp_phi} the graphical behavior of $\langle(\Delta \mathcal{\hat{p}}^{\phi})^{2}\rangle_{\textrm{ren}}^{\textrm{(j)}}$ is also given by Figure \ref{fig02_mvsm_theta_and_phi}. It is interesting to note that the asymmetry of these results, that is, $\langle(\Delta \mathcal{\hat{p}}^{\chi})^{2}\rangle_{\textrm{ren}}^{\textrm{(j)}} \neq\langle(\Delta \mathcal{\hat{p}}^{\theta})^{2}\rangle_{\textrm{ren}}^{\textrm{(j)}}=\langle(\Delta \mathcal{\hat{p}}^{\phi})^{2}\rangle_{\textrm{ren}}^{\textrm{(j)}}$, means that the homogeneity and isotropy properties associated with the Einstein's universe without boundary conditions treated in Ref. \cite{Ferreira:2023gzv} is broken when these conditions are introduced in the $\chi$ coordinates. This also bear some resemblance to the IQBM situation of a point particle in the presence of parallel planes in Minkowski spacetime. In other words, in both scenarios the expression for the velocity dispersion related to the component of the boundary condition differs from the others  \cite{camargo2018vacuum,Ferreira:2023uxs}. Once all expressions for the dispersion of the components of physical momentum have been obtained, we will now discuss the physical and mathematical aspects of these results. Essentially, the following comments consist of qualitative and quantitative analysis.

Firstly, similar to Ref.  \cite{Ferreira:2023gzv}, it is observed that for the limits $\tau_{a}\rightarrow 0$ and $a_{0}\rightarrow\infty$ the dispersion of the physical momentum will be zero. As reported in \cite{Ferreira:2023gzv},  the null result for $\tau_{a}\rightarrow 0$, which can be easily seen in the graphs of Figures \ref{fig01_mvsm_chi} and \ref{fig02_mvsm_theta_and_phi}, means that we have recovered the classical condition of null initial momentum. In fact, we must remember that this initial boundary condition was imposed in the general expression \eqref{Sec3eq04} for the particle momentum and, consequently, is also inserted in Eq. \eqref{Sec3eq06}. Therefore, this result shows the consistency of the formalism used. 
For the limit $a_{0}\rightarrow\infty$ we must first understand that it represents a physical situation in which Einstein's universe, in principle finite and compact, becomes unbounded and infinite. Therefore, this particular limit recovers Minkowski spacetime, that is, unbounded and boundaryless.

From Figures \ref{fig01_mvsm_chi}a and \ref{fig02_mvsm_theta_and_phi}a, we immediately notice that the physical dispersions for the case of Einstein's universe ($\delta^{\textrm{(E)}}=0$) are always positive and have a fixed and increasing profile. On the other hand, when boundary conditions are used, dispersions can assume positive and negative values -- see Figures \ref{fig01_mvsm_chi}b, \ref{fig01_mvsm_chi}c and \ref{fig02_mvsm_theta_and_phi}b. Furthermore, in this case there are divergences related to particular values of time $\tau_{a}^{\textrm{(E)}}=2\pi n$, with integer $n\geq 1$. These divergences, reported in reference \cite{Ferreira:2023gzv}, occur due to the compact and closed nature of Einstein's universe. Possibly, they are consequences of the classical treatment we use for the geometry of spacetime and in a more realistic scenario, in which geometry can fluctuate, perhaps the divergences will be eliminated. In this sense, for instance, in Ref. \cite{Ford:1994cr}, through a linearized gravity approach, it was shown that assuming quantum fluctuations in the geometry of spacetime can eliminate classical singularities in the light cone. 

In the case of the Einstein's universe with Dirichlet and Neumann boundary conditions, shown by Figures \ref{fig01_mvsm_chi}b, \ref{fig01_mvsm_chi}c, \ref{fig02_mvsm_theta_and_phi}b and \ref{fig02_mvsm_theta_and_phi}c, we verify that, in addition to the divergences $\tau_{a}^{\textrm{(E)}}$, there are also divergent behaviors for the time values $\tau_{a}^{\textrm{(B)}}=(2\ell+1)\pi\mp 2\chi$, with integer $\ell\geq 0$. Observing the mathematical structure of $\tau_{a}^{\textrm{(B)}}$ we note that these intermediate divergences are related to the boundary introduced in the $\chi$ coordinate. In fact, these divergences arise due to the idealized boundary conditions for Einstein's universe -- Eq. \eqref{dirichet_bc} e \eqref{neumann_bc}. We call these divergences ``intermediate'' because they occur between the integer divergences of $\tau_{a}^{(\text{E})}$, as can be seen in the plots for the physical momentum dispersion.

To conclude this section, for the sake of clarity, we would like to comment a little more on the simplification that we have used in our study, namely, the discarding of the variation of coordinates with respect to time. Considering that this is a complementary study, in the following discussions we will be more objective and recommend that the reader consult Section 3.3 of Ref. \cite{Ferreira:2023gzv} for additional details.

Firstly, it is important to mention that this simplification is an approach commonly used in the context of IQBM \cite{camargo2018vacuum,de2014quantum,yu2004vacuum,de2019remarks}. Similar to the previous study \cite{Ferreira:2023gzv}, here we also assume the small displacements condition (SDC), which tells us that the coordinates do not vary significantly with respect to time. In other words, we are considering a regime in which the possible temporal dependence of the coordinates can be neglected. In fact, in general, the coefficients $g^{i\nu}$ and the field $\varphi$ in Eq. \eqref{Sec3eq04} are both coordinate functions which, in turn, are functions of time, that is, $g^{i\nu}=g^{i\nu}(x(t))$ and $\varphi=\varphi(x(t))$. However, considering the SDC hypothesis, we used the approximation $x^{i}(t)\approx x^{i}$, which allowed us to establish Eq. \eqref{Sec3eq06} and all subsequent results. 

In view of the introduction of the simplification described above, our expressions will be subject to constraints in order to maintain the consistency and validity of the results. Intuitively, we perceive that, if the $x^{i}$ coordinates change very little with respect to time, the associated dispersion must be small, because each $x^{i}$ value will be very close of its mean value. Therefore, the dispersion assumes very small values, that is, $\langle(\Delta \hat{x}^{i})\rangle\ll 1$. In the present case, the SDC is mathematically represented by the expression
\begin{eqnarray}\label{Sec3eq11}
\dfrac{|\langle(\Delta \hat{\mathcal{z}}_{i})^{2}\rangle_{\textrm{ren}}^{(\textrm{j})}|}{a_{0}^{2}}\ll 1,
\end{eqnarray}
where $\mathcal{z}_{i}$ represents the physical length related to the $i$ component. In curved spacetime, physical ($\mathcal{z}_{i}$) and coordinate ($x^{i}$) lengths are related according to the expression $\mathcal{z}_{i}=\sqrt{|g_{ii}|}x^{i}$ \cite{adler2021general}. Note that the previous equation corresponds to a relative (dimensionless) dispersion, established by comparing the physical dispersion of coordinates with the radius of Einstein's universe, which defines a natural scale for the system. 

In light of what was exposed, a direct way to check the influences of each parameter on the SDC is to plot the relative dispersion \eqref{Sec3eq11} as a function of $\tau_a$ and observe under which circumstances the restriction will not be violated. In other words, the SDC will not be violated for time values that maintain the validity of the relation \eqref{Sec3eq11}, that is, bellow unity. In this sense, this condition will have an upper bound of validity, which corresponds to the time in which it is equal to unity. We can obtain an equation for the dispersion of coordinates by integrating Eq. \eqref{Sec3eq04} and observing that $p^{i}=m(dx^{i}/dt)$. Thus, assuming that $x^{i}(t)\approx x^{i}$, as dictated by SDC, it follows that the renormalized dispersion for the coordinates will be obtained through the relation
\begin{eqnarray}\label{Sec3eq12}
 \langle(\Delta\hat{x}^{i})\rangle_{\textrm{ren}}^{(\text{k})} = \dfrac{q^{2}}{m_{0}^{2}}\int_{0}^{\tau}dt'\int_{0}^{\tau}dt \int_{0}^{t}dt_{1}\int_{0}^{t'}dt_{2} g^{ij}_{1}g^{ij}_{2}\partial_{j_{1}}\partial_{j_{2}}\textrm{W}_\textrm{ren}^{(\text{k})}(z_{1},z_{2}),
\end{eqnarray}  
where $z_{1}=(t_{1},\chi_{1},\theta_{1},\phi_{1})$ and $z_{2}=(t_{2},\chi_{2},\theta_{2},\phi_{2})$. 
A noteworthy detail in the previous expression is the presence of the constant mass $m_{0}$ of the particle. This indicates that the introduction of SDC eliminates the dynamics of the mass from our expressions. This fact shows the importance of SDC in our approach \cite{Ferreira:2023gzv}. 

As observed in Section \ref{sec3.1}, the  mass of the particle depends on the spacetime coordinates and the $\psi$ field -- see Eqs. \eqref{Sec3eq01a} and \eqref{Sec3eq01b}. Thus, at first, $\langle(\Delta \hat{p}^{i})^{2}\rangle\neq m^{2} \langle(\Delta \hat{u}^{i})\rangle^{2}$, because in general $m$ and $u^{i}$ can depend on the field and, therefore, can possibly be correlated in some way. For this reason, all previous results were discussed considering the particle momentum. However, the implementation of SDC, as shown in Eq. \eqref{Sec3eq12}, eliminates the dynamics of the mass from our equations and thus, under this approximation regime, we can establish that $\langle(\Delta \hat{p}^{i})^{2}\rangle= m_{0}^{2} \langle(\Delta \hat{u}^{i})\rangle^{2}$. In words, in the SDC regime, the momentum dispersion corresponds to the product of the mass squared by the particle velocity dispersion. Therefore, the discussions and conclusions in the precedent sections are valid for momentum and velocity dispersion of the particle.

Following the approach described above, from Eqs. \eqref{Sec3eq12} and \eqref{Massless_GE_WF_EDNa}, in principle we can calculate the respective expression for the dispersion of coordinates and investigate the requirements to validate the SDC. Observing the structure of these expressions it is clear that this is not a trivial calculation, due to the number of operations that we must perform. In fact, the calculation and analysis of the above expression, both analytical and numerical, is a difficult task to carry out. However, careful observation allows us to circumvent these mathematical processes as well as draw a direct conclusion without any specific calculations. 

From Eq. \eqref{Sec3eq12} it is possible to observe that, in general, a decisive parameter to validate Eq. \eqref{Sec3eq11} is the ratio $\bar{q} = \frac{q}{m_{0}a_{0}}$. The factor $a_{0}$ in this expression comes from the integrand. It is noted that, this dimensionless quantity, $\bar{q}$, multiplies the entire expression resulting from integrals and derivatives in Eq. \eqref{Sec3eq12}. Then, $\bar{q}$ modulates the amplitude of the result for the dispersion of the coordinates, so that small values of $\bar{q}$ decrease the dispersion. Therefore, in summary, the smaller the values chosen for $\bar{q}$, the smaller the dispersion amplitude and the more effective the SDC hypothesis becomes. This is a conclusion that was also drawn (in detail) in the previous study \cite{Ferreira:2023gzv}. Furthermore, in similar systems considering scalar fields an analogous conclusion has already been reported in the literature \cite{Ferreira:2023uxs,de2014quantum}.

\section{Final remarkers}\label{sec4}

In this paper, continuing a previous investigation \cite{Ferreira:2023gzv}, we study the IQBM of a point particle coupled to a massless scalar field in Einstein's universe with Dirichlet and Neumann boundary conditions. In order to investigate the IQBM we calculate the dispersion in the components of the particle's physical momentum.
To perform our study, some simplifications were necessary, among them the consideration that variations in the particle position with respect to time are negligible. This hypothesis define the so-called small displacement condition (SDC). It is observed that to validate this regime it is necessary that the dimensionless ratio $\bar{q} = \frac{q}{m_{0}a_{0}}$ has sufficiently small values. Here, this conclusion was indirectly obtained through a fundamental analysis of the expression used to study the SDC.

The formalism used allowed us to write this observable, $\langle(\Delta \hat{\mathcal{p}}^{i})^{2}\rangle_{\textrm{ren}}^{\textrm{(j)}}$, in terms of Wightman functions, Eq. \eqref{Sec3eq06}, showing that these are crucial quantities for the studies carried out. Therefore, given this fact, we calculate the Wightman functions for each particular case and construct a compact expression that brings together all cases -- Eq. \eqref{Massless_GE_WF_EDNa}. This compact expression proved itself to be very useful for our purposes, since it allowed us to symbolically organize the corresponding expressions for the particle's physical momentum in a general way. This aspect can be seen in general in Eq. \eqref{Sec3eq08}, which shows that the momentum dispersion components are composed of two contributions, one related to Einstein's universe and the other to Dirichlet and Neumann boundary conditions. 

Here the results from the literature were retrieved and generalized by implementing boundary conditions in the studied model. It is noted that the introduced boundary conditions produce an asymmetry in the equations for dispersion in the components of the particle's physical momentum, thus breaking the properties of homogeneity and isotropy, reported in the previous study \cite{Ferreira:2023gzv}. We observed that the expression for the component $\chi$, Eq. \eqref{Sec3eq_physical_momentum_disp_chi}, related to the boundary condition, differs from the others, $\theta$ e $\phi$, Eqs. \eqref{Sec3eq_physical_momentum_disp_theta} and \eqref{Sec3eq_physical_momentum_disp_phi}, respectively. This is a peculiar aspect and resembles the one which occur in the IQBM of a point particle in the presence of perfectly reflecting planes in Minkowski spacetime, in where the expression for the dispersion of the particle's velocity in the direction perpendicular to the plane(s) differs from those associated with parallel directions \cite{camargo2018vacuum,Ferreira:2023uxs}. Furthermore, we observed the emergence of two sorts of divergences, namely, $\tau_{a}^{(\textrm{E})}=2\pi n$ and $\tau_{a}^{(\textrm{B})}=(2\ell+1)\pi\mp 2\chi$, with the numbers $n\geq 1$ and $\ell\geq 0$ being real and integers. The first of these,  $\tau_{a}^{(\textrm{E})}$, arises from the compact nature of Einstein's universe. On the other hand, the second, $\tau_{a}^{(\textrm{B})}$, arises due to the boundary conditions imposed on the modes of the scalar field, for the angular coordinate $\chi$, defining Einstein's universe.

{\acknowledgments}

E.J.B.F. would like to thank the Brazilian agency Coordination for the Improvement of Higher Education Personnel (CAPES) for financial support through the doctoral scholarship. The author H.F.S.M. is partially supported by the Brazilian agency National Council for Scientific and Technological Development (CNPq) under Grant No. 308049/2023-3.


\end{document}